\documentclass[structabstract]{aa}  
\pdfoutput=1
\usepackage{graphicx}
\usepackage{txfonts}
\usepackage{natbib}
\newcommand\be{\begin{equation}}
\newcommand\ee{\end{equation}}
\newcommand\bea{\begin{eqnarray}}
\newcommand\eea{\end{eqnarray}}

\begin{document}
   \title{Dust-trapping Rossby vortices in protoplanetary disks}

   \author{H. Meheut
          \inst{1}
          \and
          Z. Meliani\inst{2}          
          \and
          P. Varniere\inst{3}
          \and
          W. Benz\inst{1}
          }

   \institute{ Physikalisches Instit\"ut \& Center for Space and Habitability, Universit\"at Bern, 3012 Bern, Switzerland\\
              \email{meheut@space.unibe.ch}
		 \and
			 LUTH, Observatoire de Paris, CNRS, Universit\'e Paris Diderot, 5 place Jules Janssen, 92190 Meudon, France
         \and
             AstroParticule et Cosmologie (APC), Université Paris Diderot, 10 rue A. Domon et L. Duquet, 75205 Paris Cedex 13, France \\				
             }

   \date{\today}

 
  \abstract
   {One of the most challenging steps in planet formation theory is the one leading to the formation of planetesimals of kilometre size. A promising scenario involves the existence of vortices able to concentrate a large amount of dust and grains in their centres. Up to now this scenario has been studied mostly in 2D razor thin disks. A 3D study including, simultaneously, the formation and resulting dust concentration of the vortices with vertical settling, was still missing.} 
    {The Rossby wave instability self-consistently forms 3D vortices, which have the unique quality of presenting a large scale vertical velocity in their centre. Here we aim to study how this newly discovered effect can alter the dynamic evolution of the dust.
}
   {We performed global 3D simulations of the RWI in a radially and vertically stratified disk using the code MPI-AMRVAC.
    After the growth phase of the instability, the gas and solid phases are modelled by a bi-fluid approach, where the dust is considered as a fluid without pressure. Both the drag force of the gas on the dust and the back-reaction of the dust on the gas are included. Multiple grain sizes from $1mm$ to $5cm$ are used with a constant density distribution.}
   {We obtain in a short timescale a high concentration of the largest grains in the vortices.  Indeed, in $3$ rotations the dust-to-gas density ratio  grows from $10^{-2}$ to unity leading to a concentration of mass up to that of Mars in one vortex. The presence of the radial drift is also at the origin of a dust pile-up at the radius of the vortices. Lastly, the vertical velocity of the gas in the vortex causes the sedimentation process to be reversed,  the $mm$ size dust is lifted and higher concentrations are obtained in the upper layer than in the mid-plane.}
   {The Rossby wave instability is a promising mechanism for planetesimal formation, and the results presented here can be of particular interest in the context of future observations of protoplanetary disks.}


   \maketitle
%

\section{Introduction}

In the current formation theory, planets are supposed to be built from colliding planetesimals of kilometre or larger size, but the formation of these planetesimals is still an issue \citep{CY10}. 
 
 Due to their intermediate sizes, they cannot stick through chemical bonds and van der Vaals forces, as opposed to microscopic dust \citep{D09, BL08}. Moreover, their gravitational fields are too small to retain collision fragments \citep{B00}. 
Besides, the gas is partially supported by the radial pressure gradient and is therefore sub-keplerian. The solids in keplerian rotation feel a head-on wind which slows them down. This drag force induces a radial drift toward the central star on timescales as short as hundreds of years for meter size solids \citep{WEI77}. This timescale appears to be even shorter when compared to the planet formation timescale of a few million years.

Multiple scenarios have been proposed to overcome this difficulty.  The streaming instability \citep{YG05,JOH07} is an hydrodynamical instability that grows partially thanks to the strong coupling between gas and dust. But its domain of interest only includes regions with an increased dust-to-gas density ratio, compared with the standard value of $\rho_d/\rho_g \sim 10^{-2}$. 
Another possibility, that does not exclude the previous one, is the presence of vortices in the protoplanetary disk. 
In this scenario the meter size barrier is outstripped by the presence of vortices that can 
concentrate solids in their centre and accelerate the growth process. \citet{BAR95} have shown, with an analytical approach, that anticyclonic vortices effectively concentrate solids in their centre, and this idea was further studied by \citet{TBD96}. As this concentration effect by the vortices shares the same physics as the radial transport of solids toward the disk centre, the highest concentration is expected for the fastest drifting solids. Whereas these studies ignored the vertical structure of the disk, a 3D approach was proposed by \citet{SHE06} and \citet{HK10}. Numerical simulations of these dusty vortices have also been performed in 2D \citep{BCP99,GL99-2,GL00} and 3D \citep{JAB04} in order to investigate their ability to concentrate solids.

However these works leave unspecified the formation mechanism for the vortices and their long term evolution. 
\citet{KLA03} and \citet{KLA04} have proposed a non-linear hydrodynamical instability growing in an entropy gradient \citep{PJS07,PSJ07}, called the baroclinic instability, to form such structures.
\citet{LP10} showed that these vortices are stable structures in 3D, whereas the MHD approach of \citet{LK11} proved that this instability can form vortices only in the dead-zone. 
However the vortices migrate radially due to the radial pressure gradient \citep{PLP10} and their long term stability is not clear. 
Another proposed formation mechanism for the vortices is the Rossby wave instability (RWI). This is a linear instability that grows in the region of a pressure extremum, which avoids vortex migration \citep{MKC12}. 
This instability was first studied in 2D both analytically \citep{LOV99} and numerically \citep{LI01,VAR06}. The concentration of solids in Rossby vortices has been explored in numerical simulations \citep{INB06,LYR09}. 
Recently, \citet{RJS11} have proposed the RWI as an explanation for the non-axisymmetric submillimeter images of some transition disks. 
Moreover, an equivalent of the RWI can also exist in a magnetised disk \citep{TAV06,YL09}, extending its domain of application. 
One intriguing characteristic of these vortices is that the vertical displacements of gas in the vortices centres over the whole vertical scale height of the disk. 
This was first obtained in numerical simulations \citep{MEH10} and then confirmed analytically \citep{MYL12,L12}. 
These structures are of particular interest for the study of the dust concentration. Indeed, not only can they accelerate the concentration in the mid-plane at the centre of the vortices due to the downward flow, they also replenish the upper region of the disk with small particles thanks to the upward flow.

The goal of this paper is to investigate the behaviour of solid particles in 3D vortices formed by the RWI. 
To this end, we perform full 3D simulations of a disk subject to the RWI. 
When the vortices are formed, we follow the joint evolution of the gas and the dust, for multiple dust species, through bi-fluid simulations. 
Our paper is organised as follows. 
We first present the formation of the vortices, detailing the characteristics of the RWI, and explaining the numerical methods and the results of this gas-only simulation. 
Section \ref{sec:dust} deals with the  model we have used for the dust, which is considered as a pressure-less fluid, as well as the limits it sets. 
The resulting simulations are presented, showing the concentration of the dust in the mid-plane and its vertical stratification. 
We discuss these results in section \ref{sec:discussion}.


\section{Formation of Rossby vortices}\label{sec:formation}

 The aim of this study is to numerically follow the density of solids in Rossby vortices. 
 Here we use the term 'Rossby vortices' for vortices formed non-linearly by the RWI. 
 The first step of this study is therefore to obtain such structures through 3D numerical simulation of the RWI.

\subsection{The Rossby wave instability}

The RWI can be seen as an equivalent of the Kelvin-Helmholtz instability (see \emph{e.g.} \citealt{DRA04}) in the context of a differentially rotating disk. 
It is an inertial instability characterized by the formation of Rossby vorticity waves in the region of an inflexion point in the flow characteristics and spiral density waves propagating outward. 
The criterium for this instability is an extremum in the quantity $\mathcal L$ related to the vorticity of the equilibrium flow. 
In a non-magnetised and isentropic thin disk this quantity can be written as \citep{LI00}
\begin{equation}
\mathcal L= \frac{\Sigma \Omega}{\kappa^2}=	
\frac{\Sigma}{2(\vec\nabla\times\vec v)_z}
\end{equation}
where $\Sigma$ is the surface density, $v$ the velocity of the gas,  $\Omega$ the rotation frequency and 
$\kappa^2 = 4\Omega^2 + 2r\Omega\Omega '$ the squared epicyclic frequency 
(so that $\kappa^2/2\Omega$ is the vorticity). 
Here the prime notes a radial derivative. 
Note that a Rossby wave propagates in each gradient of $\mathcal{L}$, and the growth of the instability is related to the exchange of energy between these two waves. From this point of view, the instability can be considered as a global instability, and does not depend on the boundary conditions. 

Such an extremum can be achieved by the presence of a density bump in the disk as expected at the boundaries of the dead-zone. 
This is the region where the gas is not ionised by stellar radiation nor by cosmic rays, and turbulence driven by magneto-rotational instability is ineffective \citep{G96}.  The different viscosity rates on each side of the dead-zone is responsible for the formation of a pressure bump \citep{VAR06,LYR08,KLG09}. 
Very recently \citet{LM12} have performed the first MHD simulation of the inner edge of the dead zone in unstratified 3D showing the formation of this bump and the growth of the RWI in this region. The region of the ice-line is also expected to form an extremum in the density and entropy profile \citep{KL07} and could be a region of vortex formation, as well as the edge of planet gaps \citep{KLL03,dAD07,LJK09,YLL10,LP211}.

\subsection{Numerical methods}

To simulate the growth of the RWI in the gas disk, we use the numerical methods presented in \citet{MKC12} that we briefly summarise here. 
The governing equations are solved in cylindrical coordinates $(r, \phi, z)$, centred on the star, and read:

\begin{eqnarray}
\left\{
\begin{array}{l}
\partial_t\rho+\vec\nabla\cdot(\rho \vec v)=0\\
\partial_t \rho \vec v+\vec\nabla (\vec v\cdot\rho \vec v)+\vec\nabla p=-\rho\vec\nabla\Phi_G\\
\end{array}
\right .
\end{eqnarray}

where $\rho$ is the density of gas, $\vec v$ its velocity,
$p$ the gas pressure and $\Phi_G$ the gravitational potential of the star. There is no energy equation, the evolution of the gas is considered to be isentropic:
\be
p=S\rho ^\gamma
\ee
with the adiabatic index $\gamma= 5/3$, $S$ being a constant. The sound speed is given by
\be
c_s^2=\gamma p/\rho=S\gamma \rho^{\gamma-1}
\ee
 and the temperature by 
\be
 T=p\mu/(k\rho)=S\mu/k \rho^{\gamma-1}.
 \ee
with $\mu$ the mean molecular weight of the gas and $k$ the Boltzmann constant.
The evolution of the fluid is calculated with MPI-AMRVAC (Message Passing Interface-Adaptive Mesh Refinement Versatile Advection Code) presented in \citet{KEP11}. We used the Lax-Friedrich scheme (see \citealt{TOT96}) with a Koren limiter \citep{KOR93}.
 A global simulation of the disk is necessary to fully capture the RWI, but as the instability is symmetric about the mid-plane of the disk we only need to compute the upper half of it \citep{MEH10,MYL12}.
 For those reasons, the disk is simulated on a cylindrical grid with $r~\epsilon~ [1, 6AU]$, $\varphi~\epsilon ~[0, 2\pi]$ and $z~\epsilon~ [0, 0.5AU]$. 3 AMR levels are used which gives an increase of a factor of $8$, allowing a resolution corresponding to $(256,128,128)$ to be reached in the region of physical interest.

The initial condition is a protoplanetary disk with a pressure bump allowing the RWI to grow. The initial mid-plane density is 
\be
\rho_{z=0}=\rho_0\Big(\frac{r}{r_0}\Big)^\alpha\Big(1+\chi\exp\big(-\frac{r-r_B}{\sqrt{2}\sigma}\big)^2\Big) \,,
\ee
with $\rho_0=10^{-10}g.\,{\rm{cm}}^{-3}$ the density at $r_0=1$ AU, and $\alpha=-1.5$ the power law index of the underlying density. 
This value for the density slope gives a surface density varying approximately as $\Sigma\sim r^{-0.5}$ in the absence of the bump. 
The shape of the gaussian bump is defined by its amplitude $\chi=1$, width $\sigma=0.1$ AU and radial position $r_B=3$ AU. The vertical and radial equilibria are achieved thanks to the density and azimuthal velocity profiles:
\bea
\rho&=&\rho_{z=0}\Bigg[1-GM_*\frac{\gamma-1}{\gamma S\rho_{z=0}^{\gamma-1}}\Big(\frac{1}{r}-\frac{1}{\sqrt{r^2+z^2}}\Big)\Bigg]^{1/(\gamma-1)}\\
v_\varphi&=&\sqrt{\frac{GM_*}{r}+rS\gamma \rho_{z=0}^{\gamma-2}\partial_r\rho_{z=0}}
\eea
with $S=10^{-3}$. This gives a temperature of approximatively $ 2.5~10^2K$ at a radius of $1AU$. On top of this equilibrium state, we added small random perturbations on the radial velocity with a relative amplitude to the inner azimuthal velocity of $10^{-4}$. 
If not specified, the length is given in $AU$, the time $\tilde t$ in the code time unit corresponding to $1/(2\pi)~yr$, and the densities are normalised to $\rho_0$.

\subsection{Growth of the instability}

\begin{figure}
	\centering
   \includegraphics[width=8cm,trim=0.5cm 2cm 0cm 1cm,angle=90,clip=true]{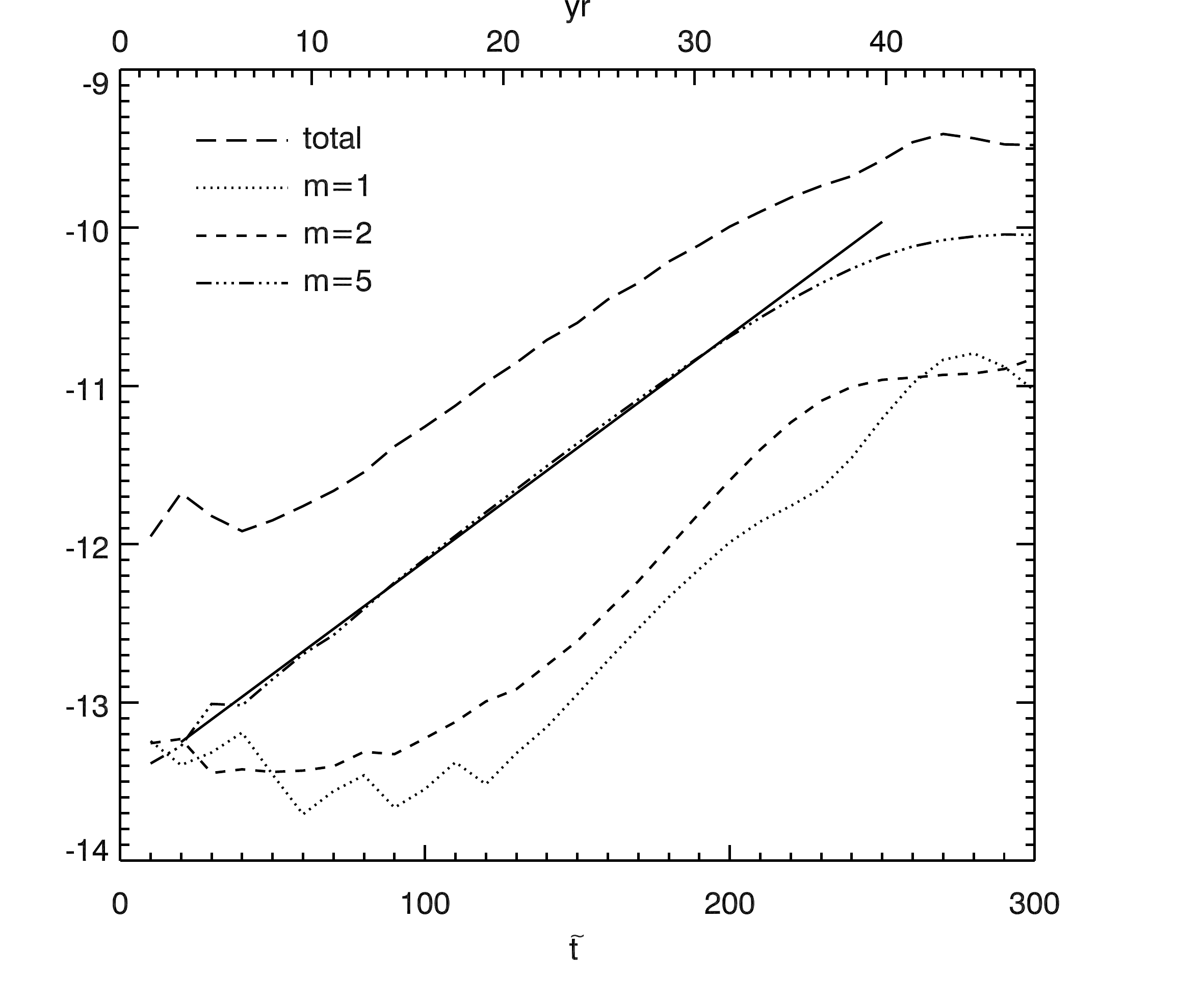}
    \caption{Amplitude of the total density perturbation and the main modes on a logarithmic scale as a function of time. The solid line is a fit of the linear growth giving a growth rate of $0.17\Omega_K(r_B)$.
              }
    \label{Fig:gasgrowth}
\end{figure}

These conditions are favourable for the RWI, and the simulation shows its growth with the formation of Rossby vortices. After the exponential growth phase that lasts for a few rotations, the instability reaches saturation as expected. At this point the amplitude of the perturbations cease to grow exponentially and maintain a nearly constant value. Fig.~\ref{Fig:gasgrowth} shows the amplitude of the density perturbations on a logarithmic scale as a function of time. At the time $\tilde t=300$ the instability has fully reached the saturation and there is no important change in the structure of the disk on a time-scale of a few rotations. This corresponds to approximately $ 50 yr$ and $14$ rotations at the radius of the bump. This evolution and the longer-term stability of Rossby vortices have been studied in the absence of dust by \citet{MKC12}, showing that the vortices survive over at least a few hundreds of years.

On the same figure, the growth of a few modes is also plotted. A mode is an element of the Fourier transform of the density: 
\begin{equation}
\rho(r,z,\varphi,\tilde t)=\sum_m \rho_m(r,z,\tilde t)\exp(-im\varphi)
\end{equation}
where the azimuthal mode number $m$ is a positive integer.
The most unstable azimuthal mode during the exponential growth is the one with $m=5$. This azimuthal mode number corresponds to the number of anticyclonic vortices, which are rotating counter to the keplerian rotation and are high-pressure regions. These five high-pressure regions will be easily identified in \mbox{Fig.~\ref{Fig:rhomid}}. In \mbox{Fig.~\ref{Fig:gasgrowth}}, the solid line corresponds to a linear fit of the amplitude growth of the density perturbation on a logarithmic scale. This fit gives a growth rate of $0.17\Omega(r_B)$ which is consistent with a linear calculation (see also the discussion in \citet{L12} on the shape of the initial bump). The growth of the instability has already been widely studied and we do not aim here to discuss it further.
At $\tilde t=300$ we add a dust population in the disk to follow its concentration in the five anticyclonic vortices as detailed in the next section. This defines the time $t=0$, when we have
\be
t=\tilde t-300/\Omega_K^{0}.
\ee

\section{Dust and gas joint evolution}\label{sec:dust}

When the vortices are self-consistently formed in the gas-only disk, we start to model the joint evolution of gas and dust.

\subsection{Bi-fluid model}

In the cylindrical coordinates the bi-fluid equations are
\begin{eqnarray}
\left\{
\begin{array}{l}
\partial_t\rho+\vec\nabla\cdot(\rho \vec v)=0\\
\partial_t \rho \vec v+\vec\nabla (\vec v\cdot\rho \vec v)+\vec\nabla p=-\rho\vec\nabla\Phi_G+\rho_d\vec f_d\\
\partial_t\rho_d+\vec\nabla\cdot(\rho_d \vec v_d)=0\\
\partial_t\rho_d\vec v_d+\vec\nabla (\vec v_d\cdot\rho \vec v_d)=-\rho_d\vec\nabla\Phi_G-\rho_d\vec f_d\\
\end{array}
\right .
\label{eq:fluid}
\end{eqnarray}

where $\rho$ and $\rho_d$ are the density of gas and dust, $v$ and $v_d$ their velocities, $p$ the gas pressure and $\Phi_G$ the gravitational potential of the central star. The drag force $\vec f_d$ between the gas and the dust is expressed in terms of the stopping time $\tau_s$
\be
\rho_d\vec f_d=\frac{\rho_d}{\tau_s}(\vec v-\vec v_d)
\label{eq:stopping}
\ee
The stopping time corresponds to the timescale of the coupling between gas and dust. A high stopping time corresponds to solids somewhat coupled to the gas, whereas the particles with \mbox{$\tau_s<<1$} will strictly follow the gas displacements.
The expression of the stopping time depends on the mean free path $\lambda$ of the particle and eventually the Reynolds number of the flow. We assume here spherical grains with radius $s_p$. The small particles with \mbox{$s_p<\frac{9}{4}\lambda$} are in the Epstein regime with \citep{TAL02}
\be
\tau_s=\sqrt{\frac{\pi}{8}}\frac{\rho_p s_p}{c_s\rho},
\ee
where the $\sqrt{\frac{\pi}{8}}$ factor comes from the expression of the mean thermal velocity in 3D \citep{TA01}, and the drag force is written as
\be
\vec \rho_d f_d=\sqrt{\frac{8}{\pi}}c_s\frac{\rho_d\rho}{\rho_p s_p}(\vec v-\vec v_d)
\ee
with $\rho_p$ the density of the individual solid particles. The internal density of a solid particle $\rho_p$ is not to be confused with the density of the dust fluid $\rho_d$.

In the following, the particle species will be defined by the non-dimensional stopping time parameter $\Omega_K^0\tau_s^0$ expressed in the mid-plane at the inner edge of the simulation. With an inner edge at $1 AU$ and a typical mass ratio of $\rho_0/\rho_p=10^{-10}$ corresponding to an individual dust particle density of $\rho_p=1g.cm^{-3}$ we have,
\be 
s_p\sim 10~\Omega_K^0\tau_s^0\rm~cm.
\ee
We will consider that each dust population corresponds to a dust size, but different stopping times could also correspond to the same size but different densities. 

The back-reaction of the dust on the gas is included in the simulation. From Eq. \ref{eq:stopping}, one can define the gas stopping time $\tau_{s,g}$ as in $\rho_d\vec f_d=\frac{\rho}{\tau_{s,g}}(\vec v-\vec v_d)$ so $\tau_{s,g}=\frac{\rho}{\rho_d}\tau_s$. This time-scale for the back-reaction decreases with increasing dust density. When the dust-to-gas ratio $\rho_d/\rho$ is small, the dynamics is dominated by the gas.
This is not the case anymore when the dust density is of the order of the gas density and the back-reaction of the dust on the gas cannot be neglected. Initially the back-reaction is negligible but will become more and more important when the dust is concentrated in the vortices and approaches the gas density.

Extensive tests of the multi-fluid module of AMRVAC will be soon published, and first tests have already been presented in \citet{vMK11}.

\begin{table}
\begin{center}
\begin{tabular}{cccc}
population&$\Omega_K^0\tau_s^0$&$\Omega_K^{r_B}\tau_s^{r_B}$&s(cm)\\
\hline
1&0.010 & 0.009&0.1\\
2&0.020 & 0.017&0.2\\
3&0.030 & 0.026&0.3\\
4&0.050 & 0.042&0.5\\
5&0.100 & 0.085&1\\
6&0.200 & 0.17&2\\
7&0.300 & 0.256&3\\
8&0.500 & 0.427&5\\
\label{Tab:dustpop}
\end{tabular}
\caption{Characteristics of the eight dust populations. The stopping time is defined at the inner edge of the disk and we also give its initial value at the radius of the bump. The last column gives an estimation of the dust size for each population.}
\end{center}
\end{table}

\subsection{Parameters}

We have performed eight simulations with different populations of dust presented in Table \ref{Tab:dustpop}. The range of dust sizes corresponds to the intermediate regime where the dust and the gas are partially coupled. The dust is added to the gas disk with a density of $1\%$ of the initial gas density.
The dust density is initially axisymmetric with a purely keplerian velocity.

The simulation was run over $16yr$ corresponding to $3$ keplerian rotations at the position of the vortices. The rotation period at the bump radius $T_B=2\pi\Omega_K^{-1}(r_B)$ will be used as a time unit.

\subsection{Validity of the model}

\begin{figure}
	\centering
	 \includegraphics[width=\linewidth]{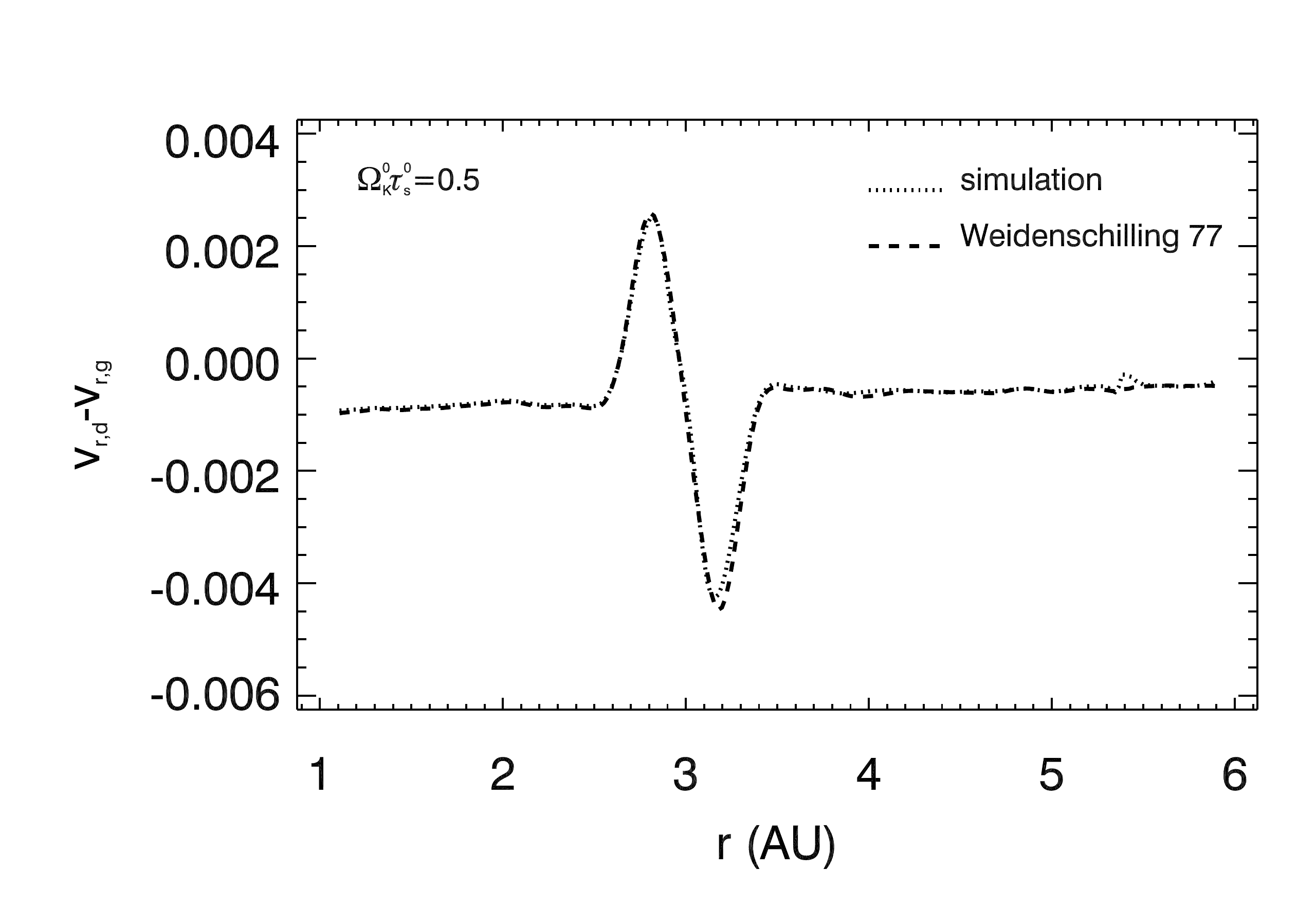}
     \includegraphics[width=\linewidth]{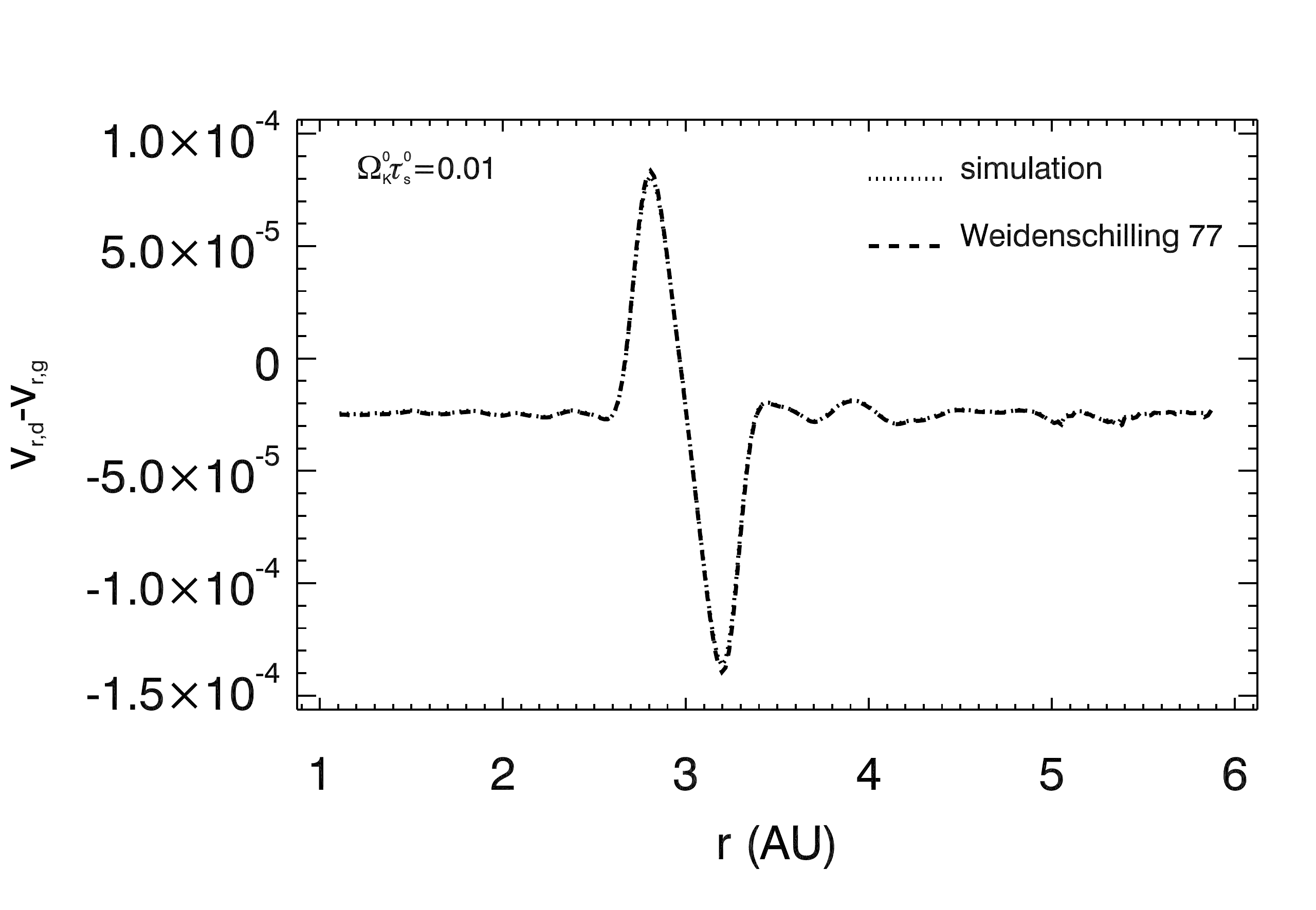}
    \caption{Radial velocity difference between dust and gas obtained in the simulation in a dotted line and with the approach of \citet{WEI77} in a dashed line. The upper and lower plots are obtained for $\Omega_K^0\tau_s^0=0.5$ and $0.01$. See text for details.
              }   
    \label{Fig:weiden}          
\end{figure}

\citet{Her09} has shown that the bi-fluid approximation where the dust is modelled as a pressureless fluid, is only valid for an adimensional stopping time $\Omega_K\tau_s<0.5$ due to the low velocity dispersion. This is the maximum value chosen at the inner edge of the disk, so the pressureless approximation is valid.

The model of the dust as a continuous fluid is valid if there are enough solid particles in each grid cell and enough collisions to define a mean state. These conditions are fulfilled as the Stokes number $\Omega_K\tau_s<1$. 

The Epstein regime corresponds to particles sizes $s_p<\frac{9}{4}\lambda$, where $\lambda$ is the mean free path of the gas molecule. This condition is directly related to the Knudsen number $Kn=\lambda/s_p$. As the main constituent of the gas is molecular hydrogen, the mean free path is written
\bea
\lambda&=&\frac{\mu}{\rho \sigma_{H_2}}
\eea 
where $\mu=3.9\times10^{-24}g$ is the mean molecular weight of a 5:1 $H_2$-$He$ mixture, and $\sigma_{H_2} = 2\times10^{-15} cm^2$ is the cross section of molecular hydrogen. This gives a maximum grain size of $~45 cm$ at $1AU$. The mean free path increases with distance to the central star as the gas density decreases. As we consider here the region of the disk around $3AU$, the Epstein regime is valid for the grains we consider, with the maximum size around $5cm$. We also note that the Reynolds number is
\be
Re=\frac{s_p\Delta v}{\lambda c_s}<1,
\ee
 which is consistent with the Epstein regime.

\section{Results}\label{sec:results}

\subsection{Test of the dust radial drift}\label{sec:weiden}

To test the numerical method in the context of protoplanetary disks, the results of the simulations have been compared with the approach of \citet{WEI77}. In the stationary limit, the difference between the radial velocity of the dust and the gas is given by
\be
v_{r,d}-v_{r,g}=\frac{\Delta g}{\Omega_K^2\tau_s+1/\tau_s},
\label{eq:weiden}
\ee
with\be
\Delta g=\frac{1}{\rho}\partial_r p.
\ee
The resulting radial velocities at the end of the simulations are plotted in \mbox{Fig.~\ref{Fig:weiden}} for populations $1$ and $8$ and compared to Eq. (\ref{eq:weiden}). Here all the quantities have been azimuthally averaged to soften the dynamical effects of the spiral density waves and vortices. There is a good agreement between the two approaches for all dust sizes, with a relative difference of about $3\%$.

\subsection{Time evolution}

\begin{figure*}
	\centering
   \includegraphics[height=11.5cm,trim=3.2cm 1cm 1.cm 0cm,clip=true]{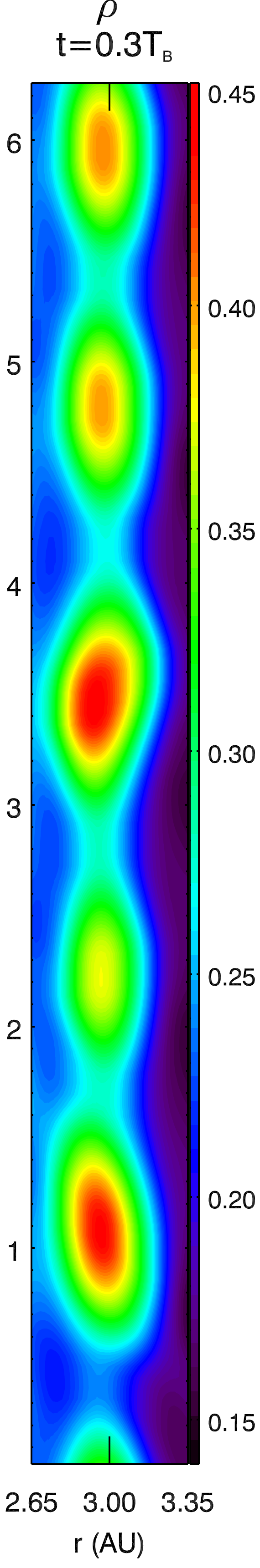} 
   \includegraphics[height=11.5cm,trim=3.35cm 1cm 0.7cm 0cm,clip=true]{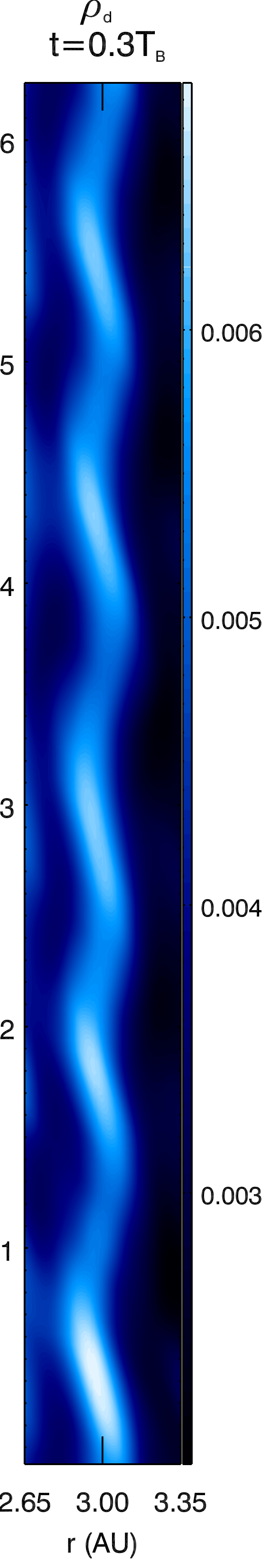} 
   \includegraphics[height=11.5cm,trim=3.9cm 1cm 1.5cm 0cm,clip=true]{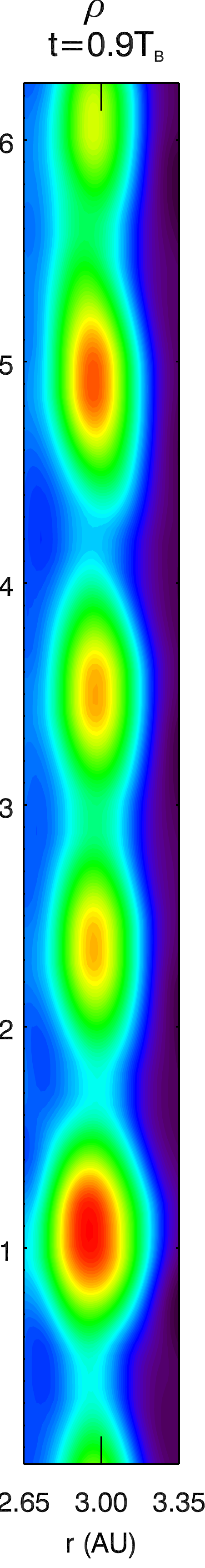} 
   \includegraphics[height=11.5cm,trim=3.35cm 1cm 0.7cm 0cm,clip=true]{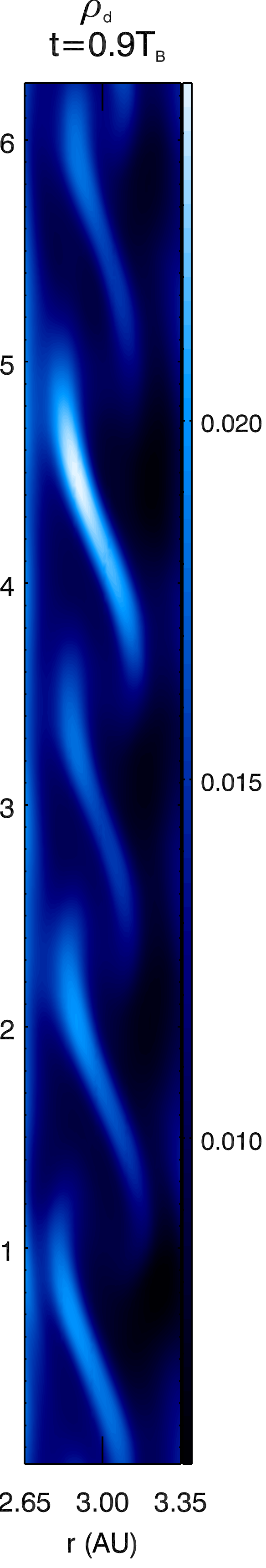}      
   \includegraphics[height=11.5cm,trim=3.9cm 1cm 1.5cm 0cm,clip=true]{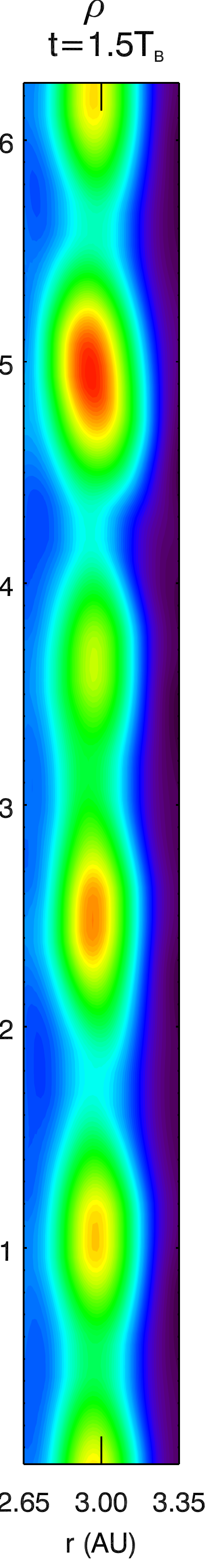}
   \includegraphics[height=11.5cm,trim=3.35cm 1cm 0.7cm 0cm,clip=true]{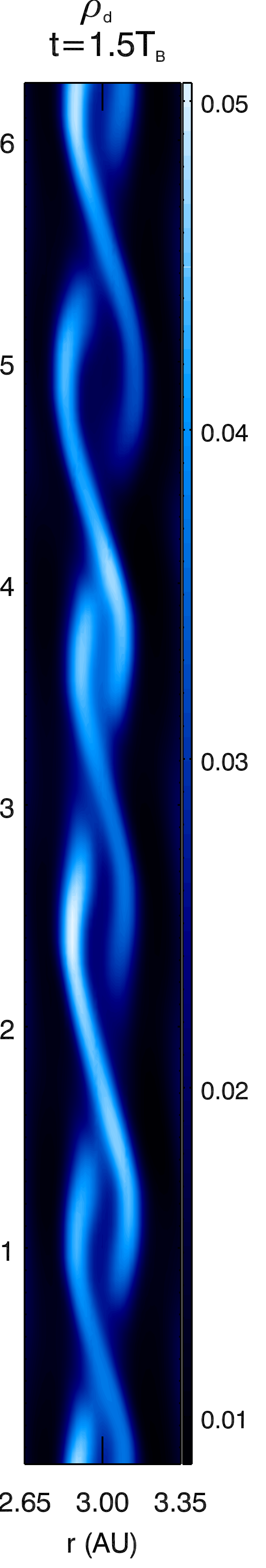}   
   \includegraphics[height=11.5cm,trim=3.9cm 1cm 1.5cm 0cm,clip=true]{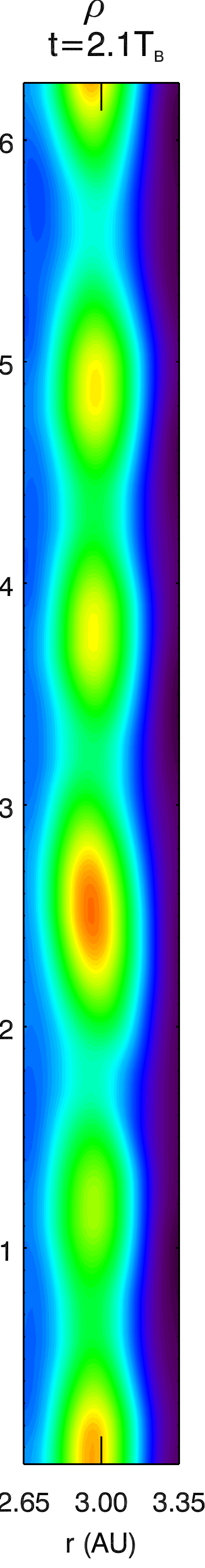}
   \includegraphics[height=11.5cm,trim=3.35cm 1cm 0.7cm 0cm,clip=true]{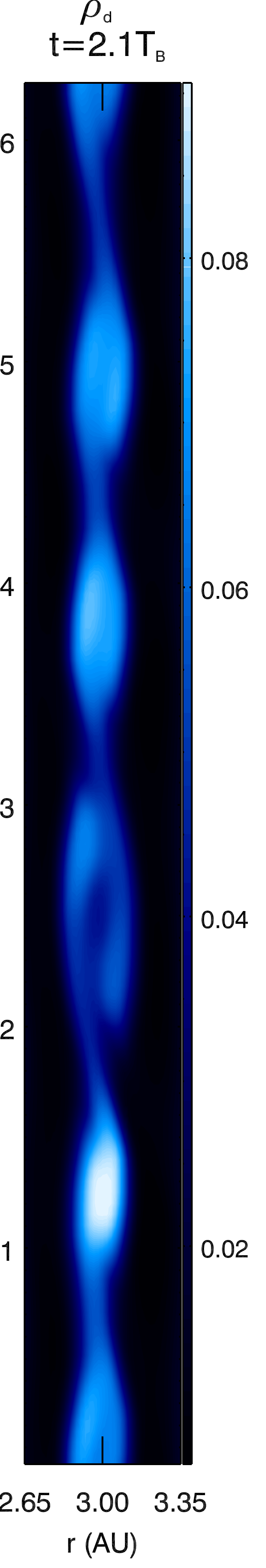} 
   \includegraphics[height=11.5cm,trim=3.9cm 1cm 1.5cm 0cm,clip=true]{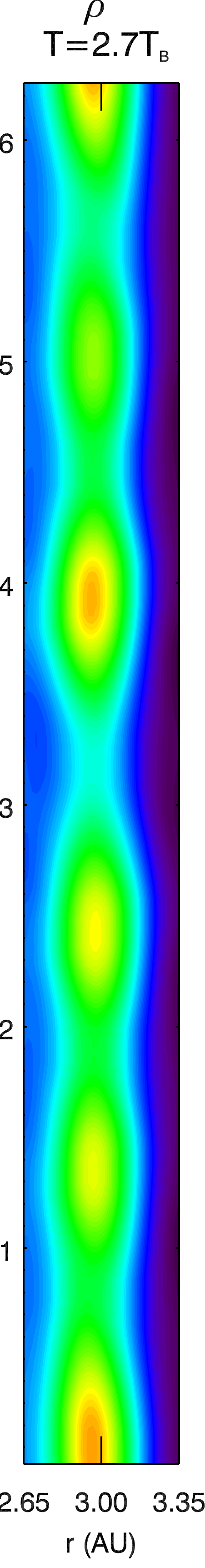}
   \includegraphics[height=11.5cm,trim=3.35cm 1cm 0.7cm 0cm,clip=true]{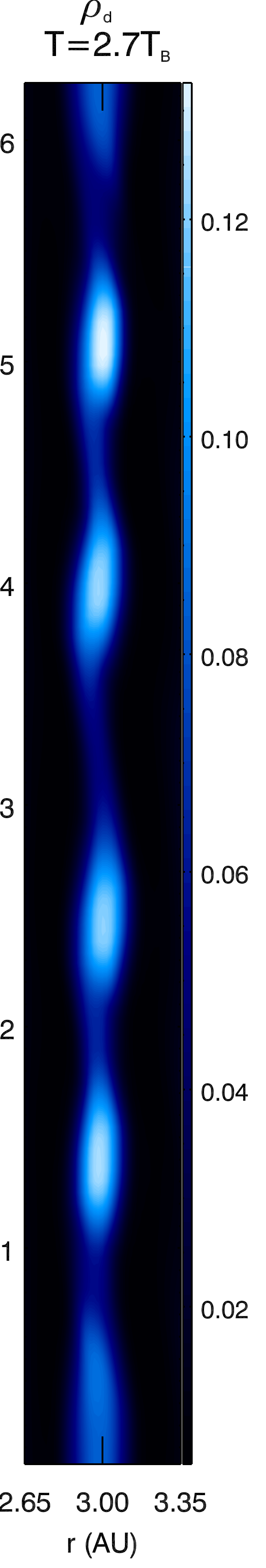}  
    \caption{Evolution of the mid-plane density of gas and $2cm$ dust ($\Omega_K^0\tau_s^0=0.2$). The times are given in keplerian time $T_B$ at $r=r_B=3AU$ after the addition of the dust grains. Note that $r_B$ is the position of the initial density bump. At $t=0$ the dust population is axisymmetric. The densities are normalised to $\rho_0$ and the colorbar used for the dust density follows the increase of this quantity whereas the gas density colorbar is fixed.}
    \label{Fig:evol}
\end{figure*}

\begin{figure}
	\centering
   \includegraphics[height=\linewidth,trim=0.5cm 1cm 0.7cm 1.4cm,clip=true,angle=90]{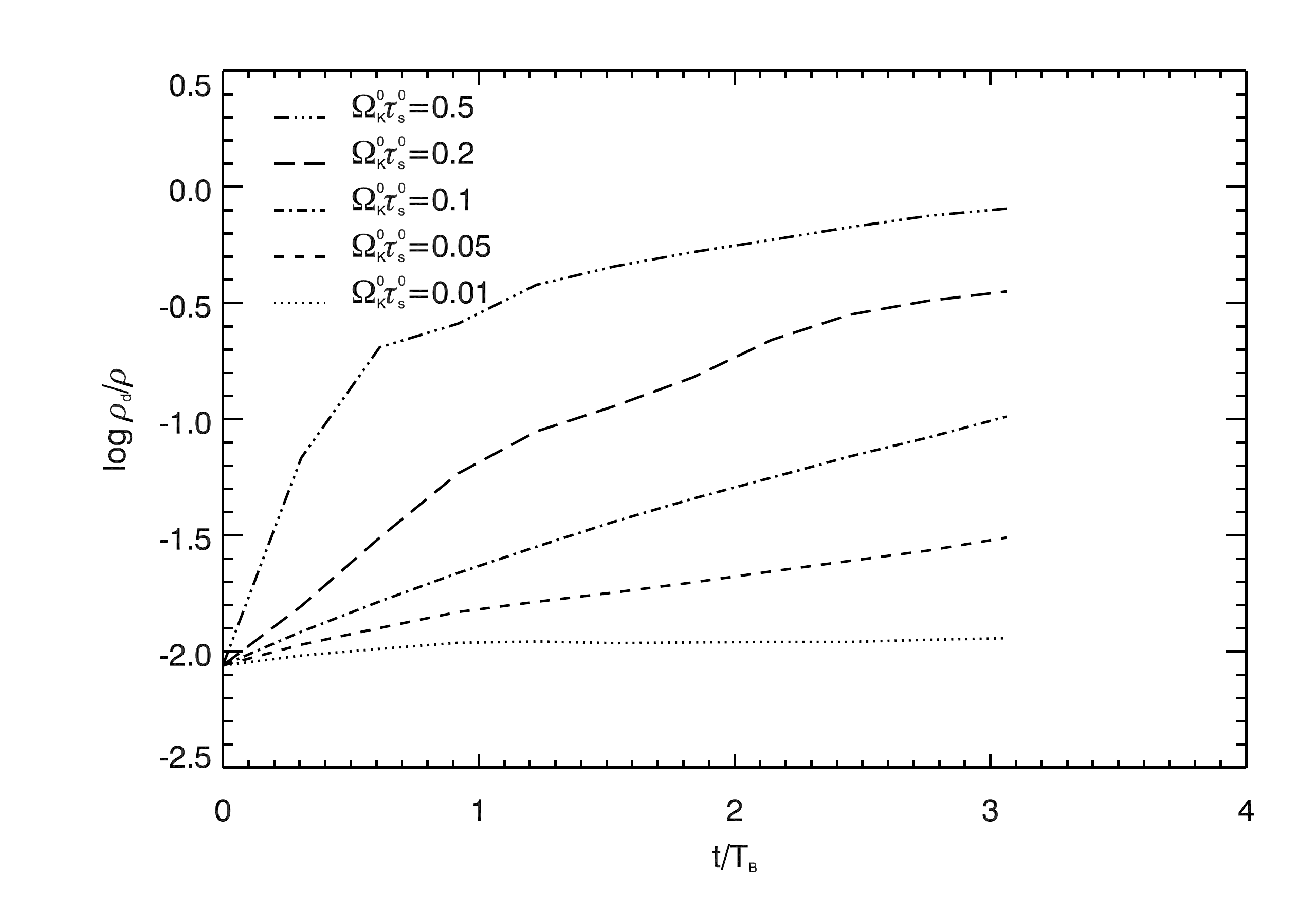}
    \caption{Maximum dust density relative to maximum gas density in the mid-plane in a logarithmic scale as a function of time for different solid sizes.}
    \label{Fig:timedust}
                  
\end{figure}

The time evolution of the gas and dust density over $3$ keplerian rotations are plotted in \mbox{Fig.~\ref{Fig:evol}} for grains of intermediate size ($2~cm$). As the gas evolves, the vortices rotate during the simulation with the frequency given by the characteristics of the RWI. The dust follows this azimuthal displacement while its concentration is modified by the presence of the vortices. 
Whereas the initial dust density is axisymetric, the vortices first seems to expel the dust from their centre: at $t=1.5T_B$ the density of dust is then higher in the surroundings of the anticyclonic vortices than in their centres. 
This is a transitory phase as the primary effect of the vortices is to induce rotation around their centre. The length of the transitory phase is related to the stopping time ($\tau_s^{r_B}=0.22T_B$ here) with a longer transitory phase for shorter stopping time when the dust is better coupled to the gas (in the limit of small solids: $\Omega_K\tau_s\le 1$). 
The dust is then concentrated directly inside the high-pressure anticyclonic vortices. See also \mbox{Fig.~\ref{Fig:stream}} showing the dust velocity streamlines which is discussed in the next section. The decrease of the gas density in the vortices when the dust density reaches high values is discussed in paragraph \ref{sec:back}. 

In \mbox{Fig.~\ref{Fig:evol}} the colour bar used for the dust density changes from one plot to another. The reason is the constant increase in dust density. This is confirmed in \mbox{Fig.~\ref{Fig:timedust}}, where the maximum value of the mid-plane dust density is plotted as a function of time. We recall that during the transitory phase, this maximum is not always reached at the centre of the vortices as can be seen in \mbox{Fig.~\ref{Fig:evol}}. In the first phase, the vortices collect the solids that are in their surroundings leading to a very fast increase of dust density. Then, for the populations that have reached the quasi-permanent state at the end of the simulation, namely those with $\Omega_K^0\tau_s^0\ge0.2$, the dust density continues to increase due to vertical settling and radial drift. This is also visible in \mbox{Fig.~\ref{Fig:rhodmoy}} which will be presented in the next section.

\subsection{3D dust concentration}

\begin{figure*}
	\centering
   \includegraphics[height=12cm,trim=2cm 1cm 0.5cm 0cm,clip=true]{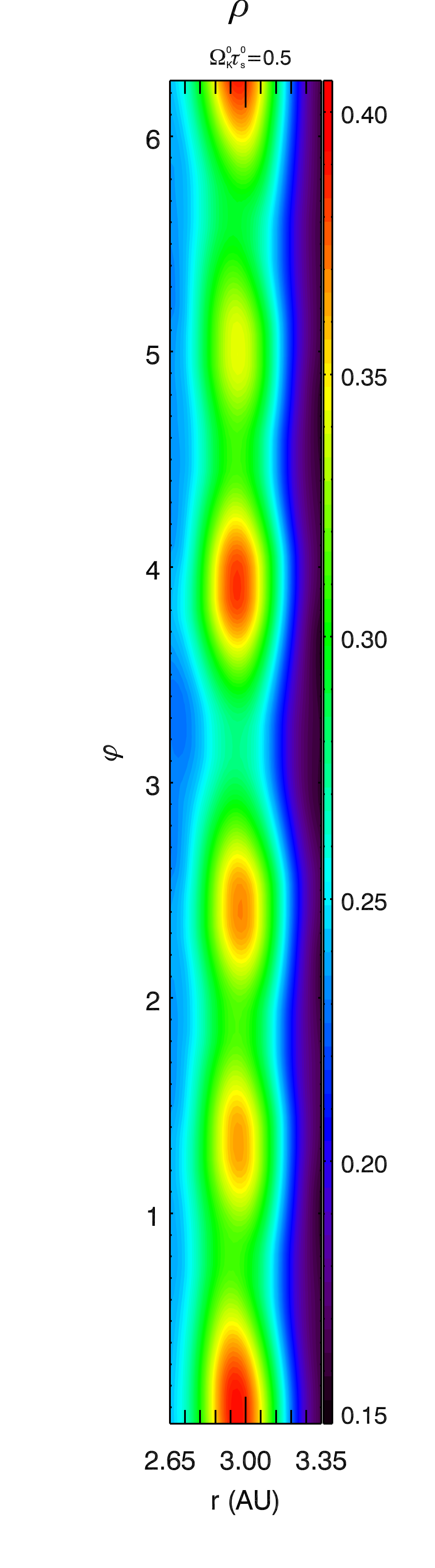}
   \includegraphics[height=12cm,trim=3.2cm 1cm 0.5cm 0cm,clip=true]{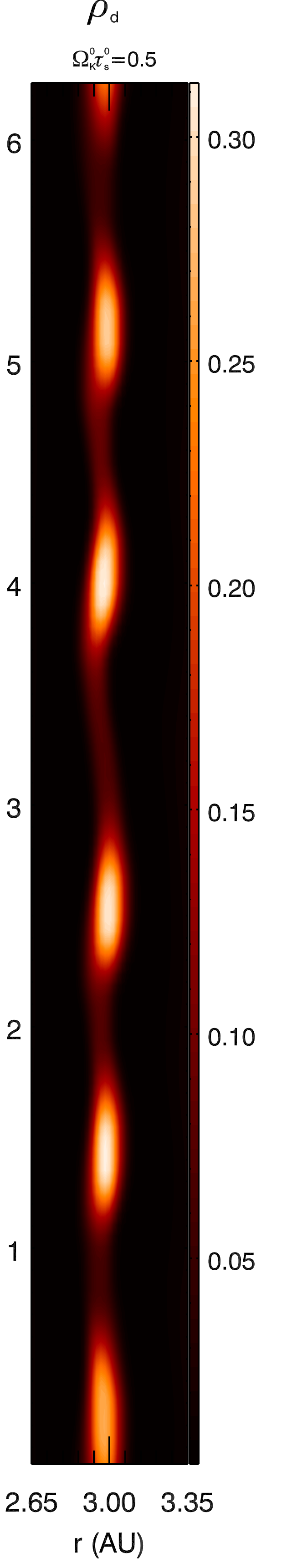}
    \includegraphics[height=12cm,trim=3.2cm 1cm 0.5cm 0cm,clip=true]{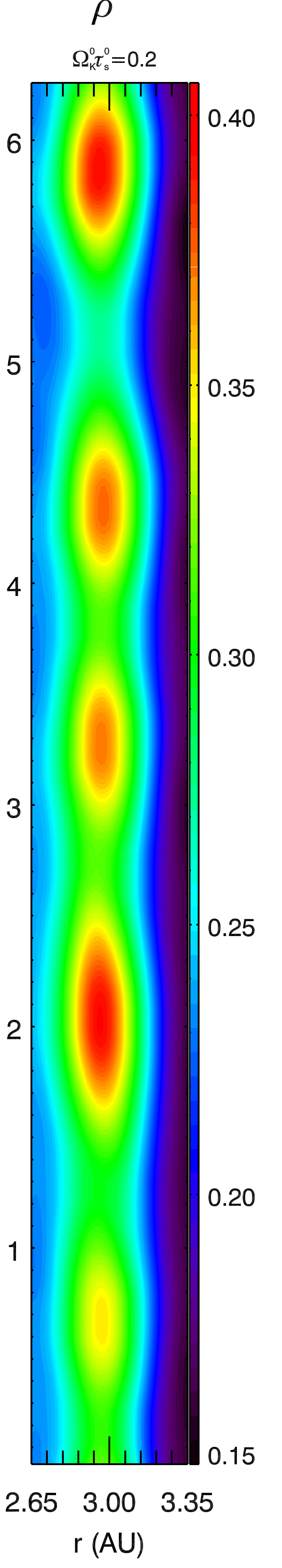}  
   \includegraphics[height=12cm,trim=3.2cm 1cm 0.5cm 0cm,clip=true]{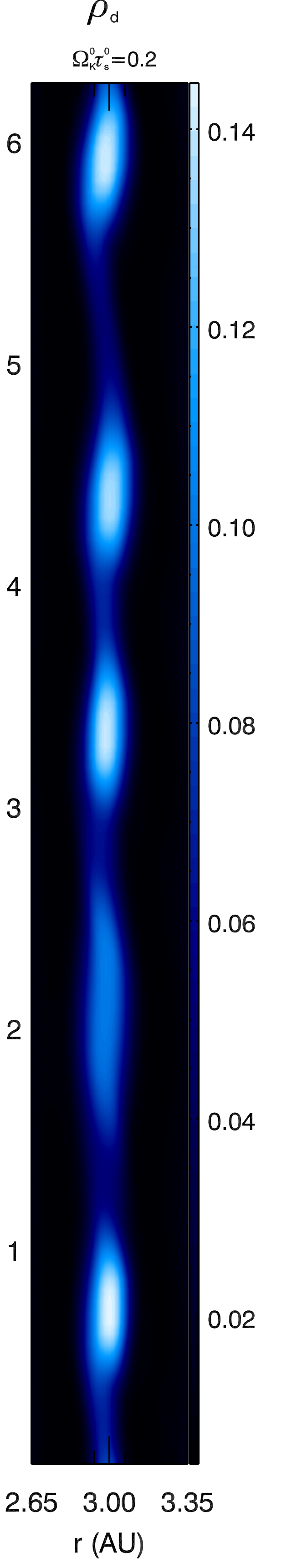}
   \includegraphics[height=12cm,trim=3.2cm 1cm 0.5cm 0cm,clip=true]{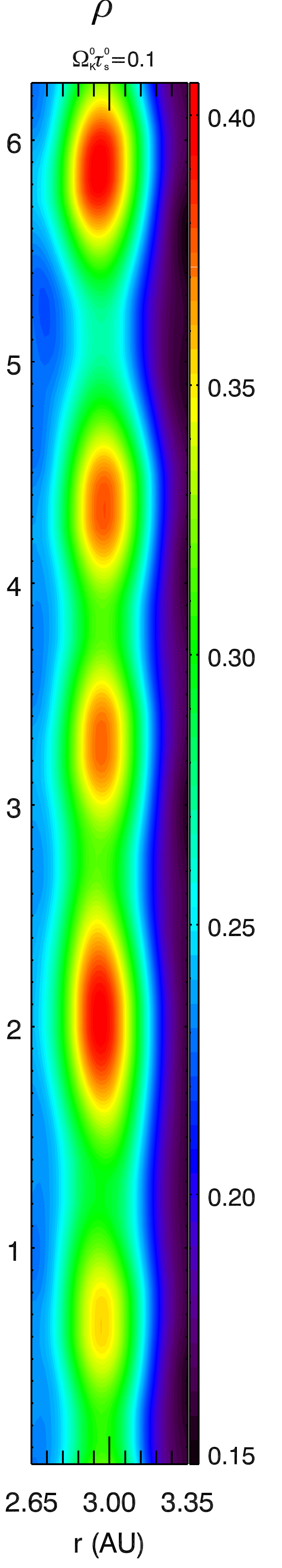}  
   \includegraphics[height=12cm,trim=3.2cm 1cm 0.5cm 0cm,clip=true]{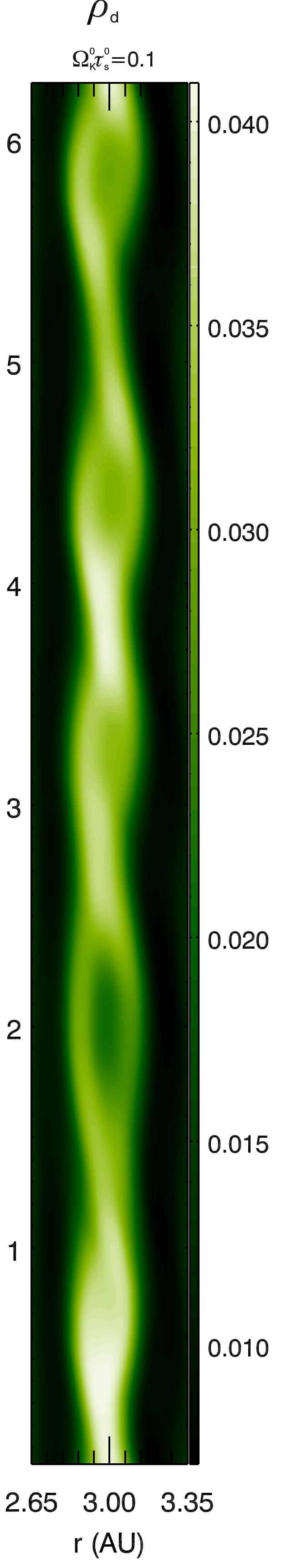}
   \includegraphics[height=12cm,trim=3.2cm 1cm 0.5cm 0cm,clip=true]{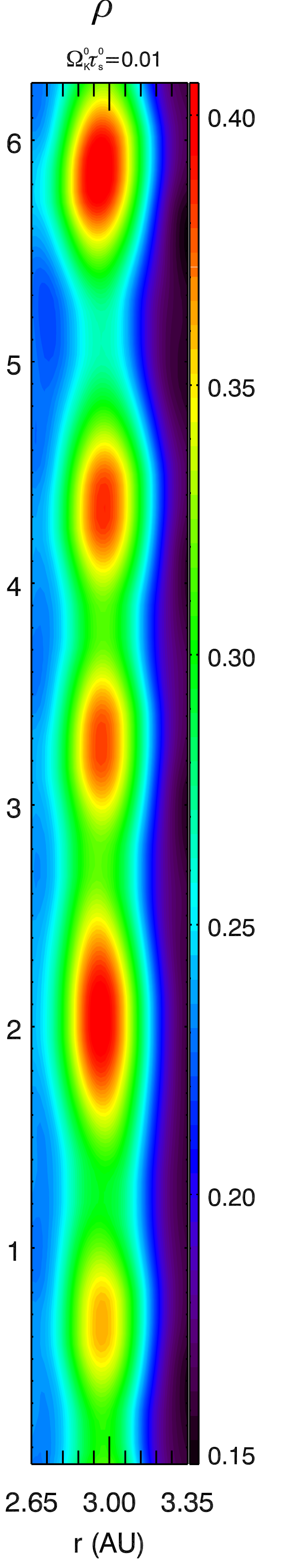}   
   \includegraphics[height=12cm,trim=3.2cm 1cm 0.5cm 0cm,clip=true]{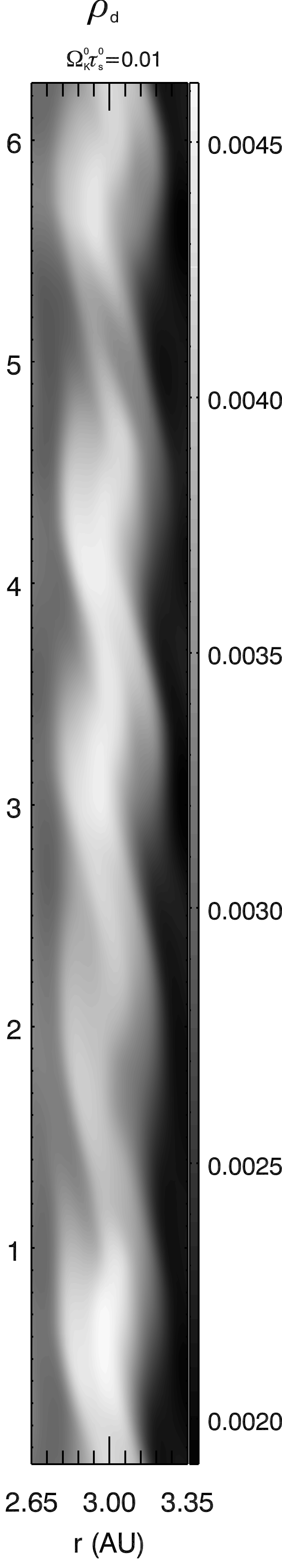}  
    \caption{Mid-plane density of gas and dust for $\Omega_K^0\tau_s^0=0.5,0.2,0.1$ and $0.01$ after $3$ rotations. As the range in density vary widely between populations, different colour tables are used to avoid misunderstandings.}
    \label{Fig:rhomid}
\end{figure*}

\begin{figure}
	\centering
   \includegraphics[height=15cm,trim=1cm 1cm 0.cm 0cm,clip=true]{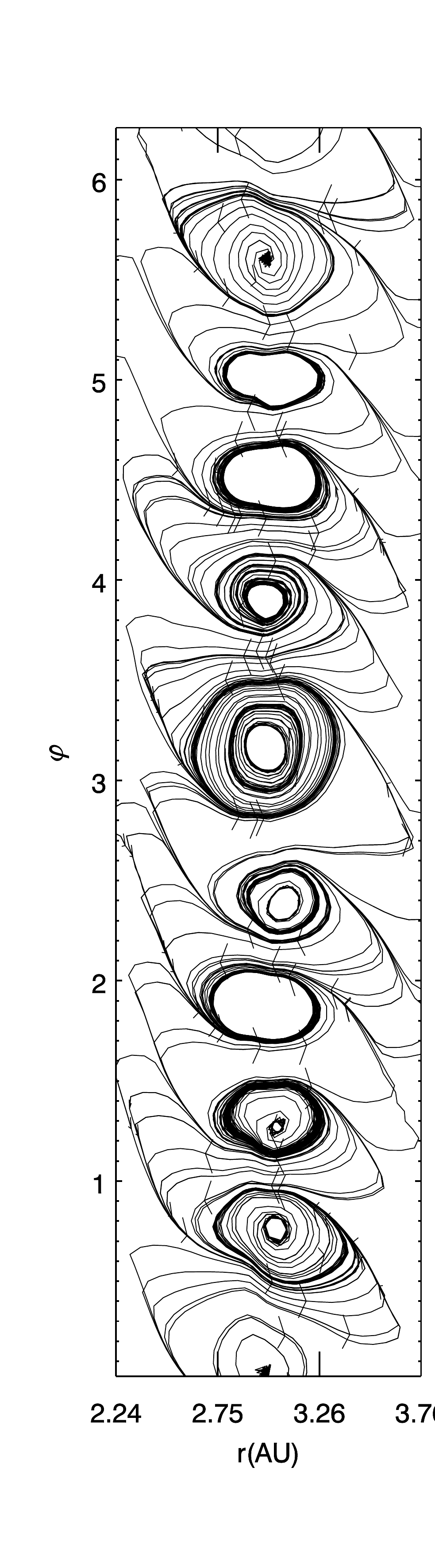}
   \includegraphics[height=15cm,trim=1cm 1cm 0.cm 0cm.clip=true]{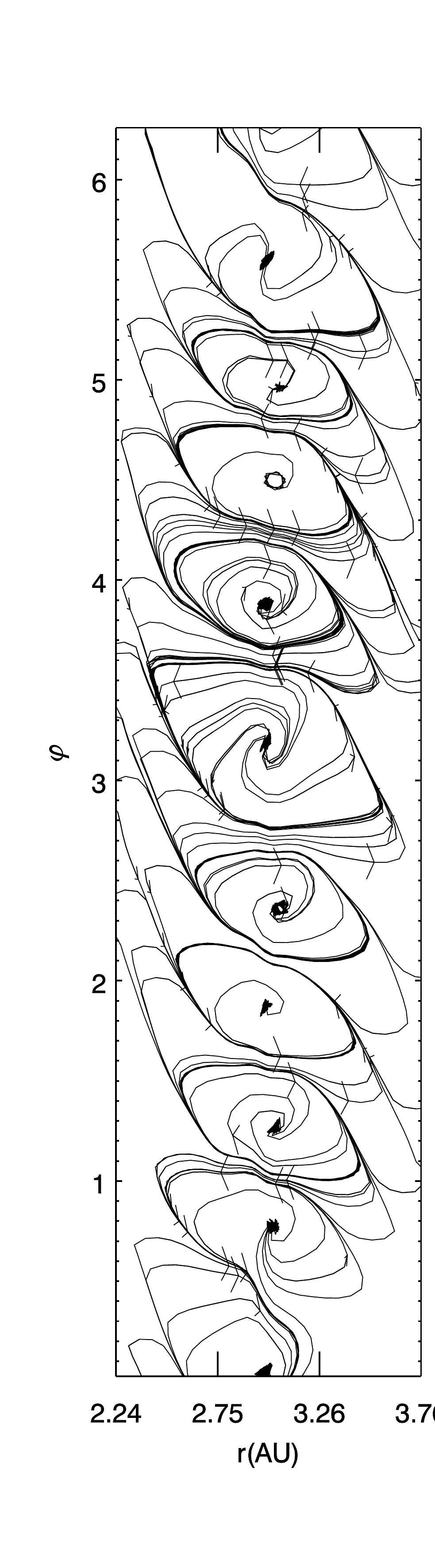}   
    \caption{Perturbed velocity streamlines of gas (\emph{left}) and $5cm$ grains (\emph{right}) in a rotating frame.
              }   
    \label{Fig:stream}          
\end{figure}

\begin{figure}
	\centering
   \includegraphics[height=9cm,trim=0.8cm 1cm 1.2cm 1.6cm,clip=true,angle=90]{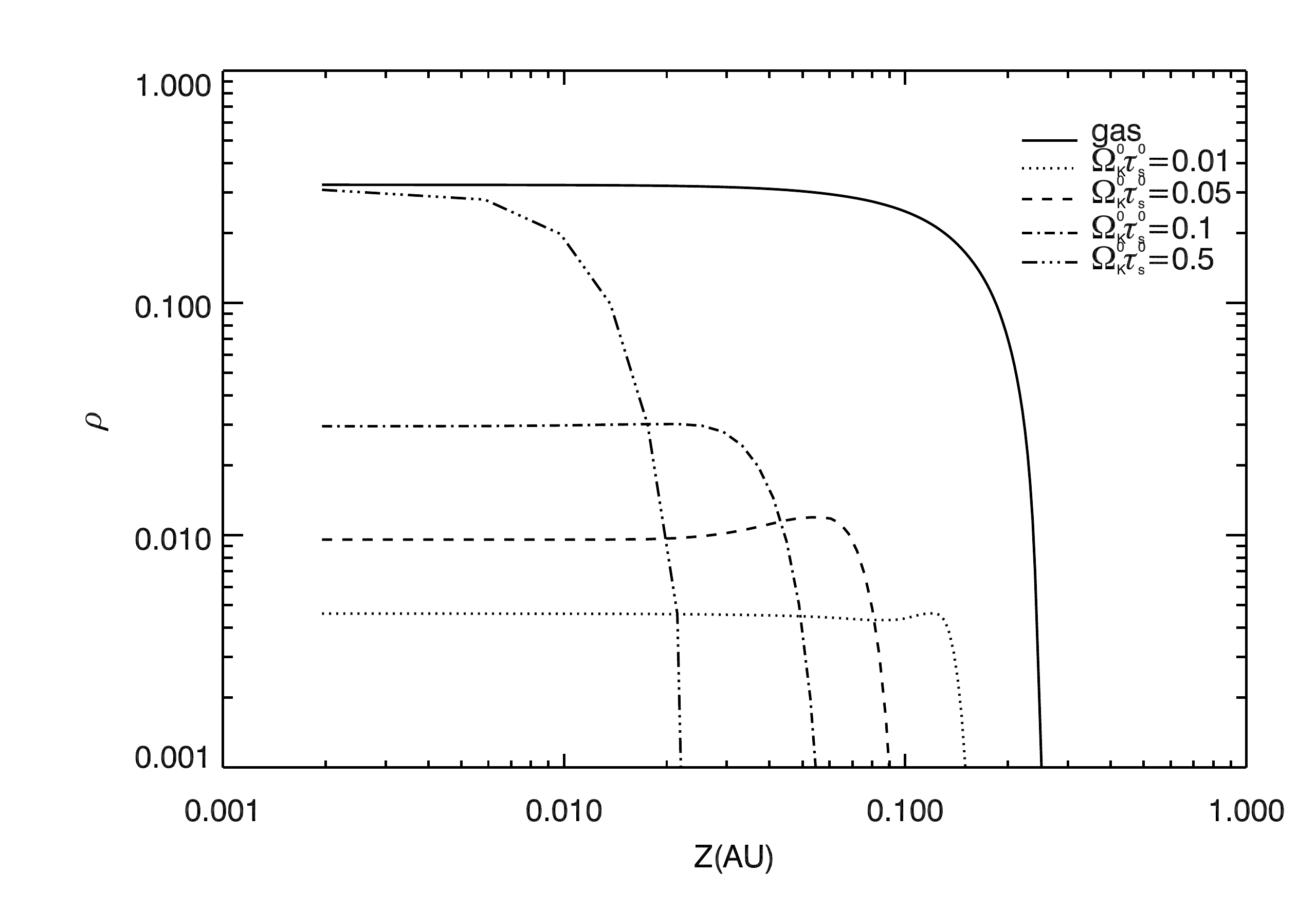}
    \caption{Vertical profile of dust and gas densities in an anticyclonic vortex on logarithmic scales.
    \label{Fig:vert}
              }    
\end{figure}

The anticyclonic Rossby vortices effectively concentrate the solid particles with an efficiency depending on the dust size. 
\mbox{Fig.~\ref{Fig:rhomid}} shows the mid-plane densities of gas and dust at the end of selected simulations. 
The colorscale is the same for the gas density obtained in the different simulations but each dust population is plotted with a different colorscale as the values differ largely from one grain size to another. 
The $1mm$ dust grains do not show a larger concentration inside the vortices even if some structures do appear. 
There is then a minimum size for the dust to be concentrated inside vortices on a rotation timescale. 
Due to the higher gas density in the anticyclonic vortices, the dust-to-gas ratio is even lower in these regions. 
In the mid-plane, the highest value of $\rho_d/\rho$ is $0.015$, corresponding to a small increase of $50\%$ (\mbox{Fig.~\ref{Fig:timedust}}). 
Indeed the smaller grains are well coupled to the gas and follow the gas streamlines plotted in \mbox{Fig.~\ref{Fig:stream}}. 
For larger particles, two different behaviours are observed at the end of the simulation. 
The $1cm$ and smaller grains are in the transitory phase, as were the $5cm$ grains at $t=2.1T_B $ (\mbox{Fig.~\ref{Fig:evol}}).
During this transitory phase, the dust density is lower in the centre of the vortices than in the surroundings. 
For the larger solids, such as the $2cm$ population, the `stationary' phase is reached with the highest density in the centre of the anticyclonic vortices and a dust-to-gas ratio $\rho_d/\rho\sim 0.4$. 
For populations up to $3cm$ the drag force of the dust on the gas is negligible and no important difference can be seen on the gas. 
The largest grains ($5cm$) are highly concentrated in the five anticyclonic vortices reaching a dust-to-gas ratio $\rho_d/\rho$ of the order of unity in the mid-plane (see also \mbox{Fig.~\ref{Fig:vert}}). 

\begin{figure}
	\centering
   \includegraphics[height=\linewidth,angle=90,trim=0.5cm 0.8cm 1cm 1.4cm,clip=true]{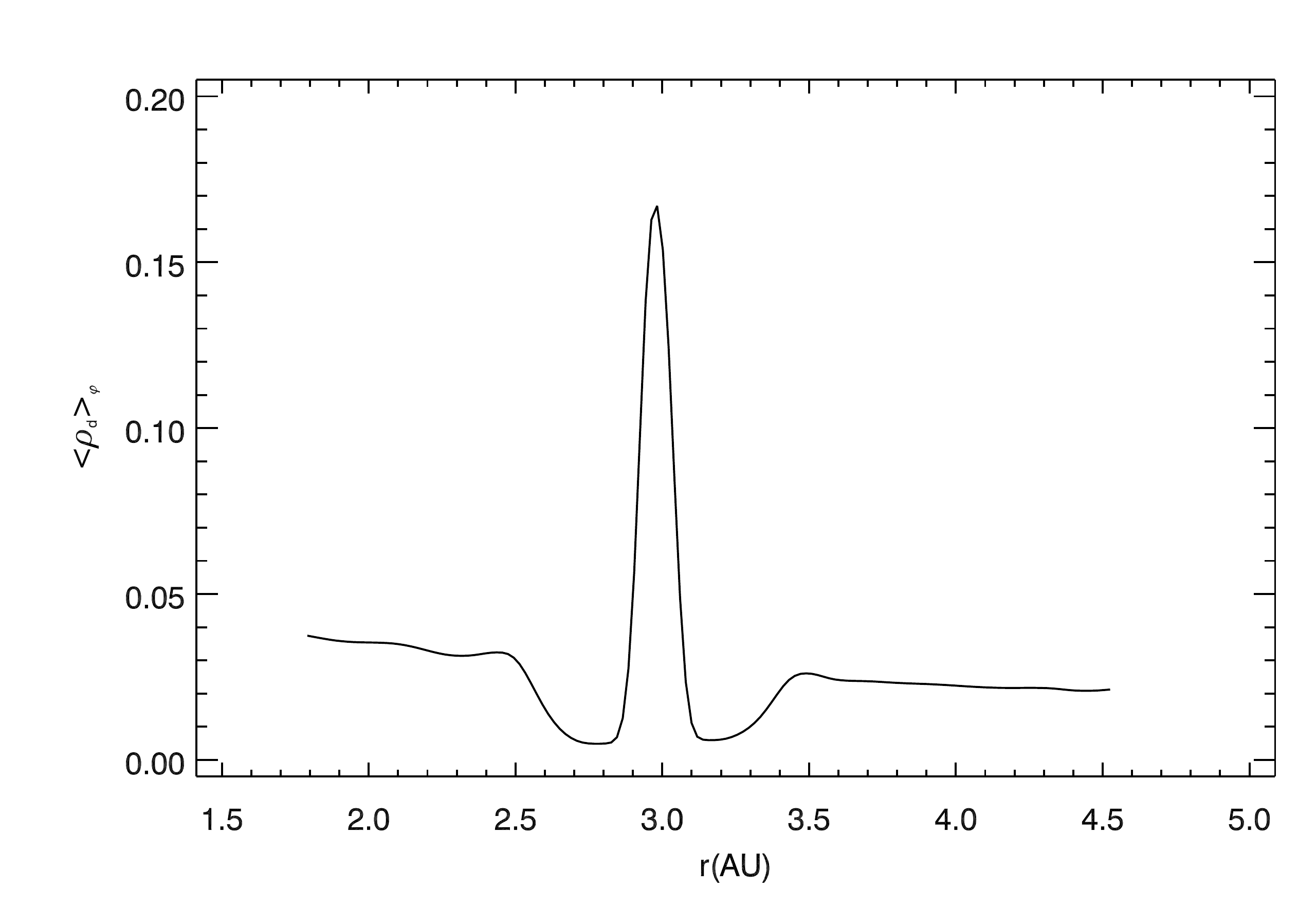} 
    \caption{Mean mid-plane density ($10^{-10}g.cm^{-3}$) of the $\Omega_K^0\tau_s^0=0.5$ dust population at the end of the simulation.
              }  
   \label{Fig:rhodmoy}           
\end{figure}

This fast concentration of dust is confirmed by the dust streamlines.
The shape of the velocity streamlines are shown in \mbox{Fig.~\ref{Fig:stream}} for both the gas and the $5cm$ dust. 
The streamlines are calculated from the velocity field with a second order Runge-Kutta integrator. 
No significative change is observed on the shape of the vortex streamlines of the gas when the dust is present. 
For both dust and gas the anticyclonic vortices have converging streamlines (corresponding to an upward flow), whereas the cyclonic vortices have diverging streamlines (corresponding to a downward flow). 
The streamlines corresponds either to a limit cycle around the vortex or to a direct displacement toward the centre in anticyclonic regions, the opposite is observed in cyclonic regions which expel the dust.
It is interesting to note that the width of the dust vortex-like structures is larger than those of the gas. 
And the dust that reaches the border of these structures directly falls to the centre. 
The dust concentrated in the vortices was initially in a zone larger than the width of the vortices. 
The 'feeding zone' has an approximate size of about ten times  the width of the initial gaussian density bump, which corresponds to about $1AU$ in our case.
\mbox{Fig.~\ref{Fig:rhodmoy}} shows the resulting mid-plane density profile averaged on the azimuthal direction, for the $5cm$ grain population. 
This profile is characterised by a very high bump at the radius of the vortices but also a depletion in the surroundings. 
This depletion is not exactly symmetric toward the centre of the vortices, with a larger mass reduction in the inner region ($r<r_B$). 
This is due to the global radial drift of the dust which refills the outward region whereas the vortices forbid inward migration of dust. 
There is then a dust pile-up at the radius of the vortices. 

\begin{figure*}
	\centering
   \includegraphics[height=0.9\textwidth,trim=2.6cm 0cm 3.25cm 2cm,clip=true,angle=90]{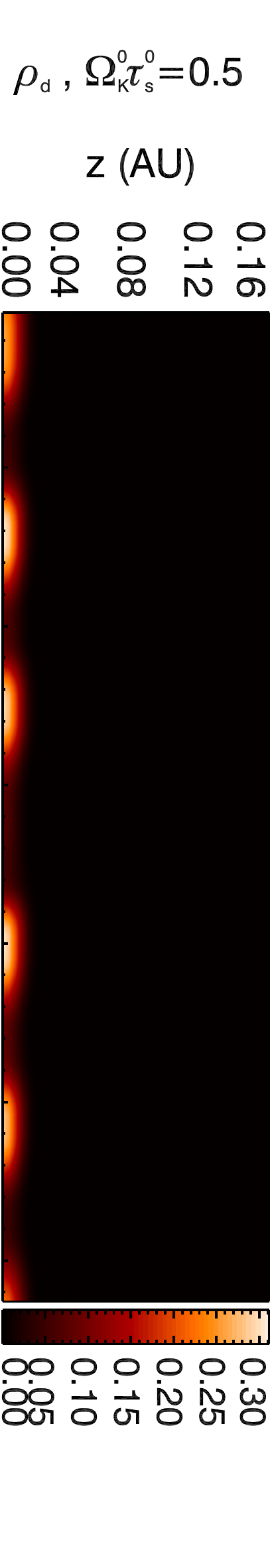} 
   \includegraphics[height=0.9\textwidth,trim=2.6cm 0cm 3.2cm 2cm,clip=true,angle=90]{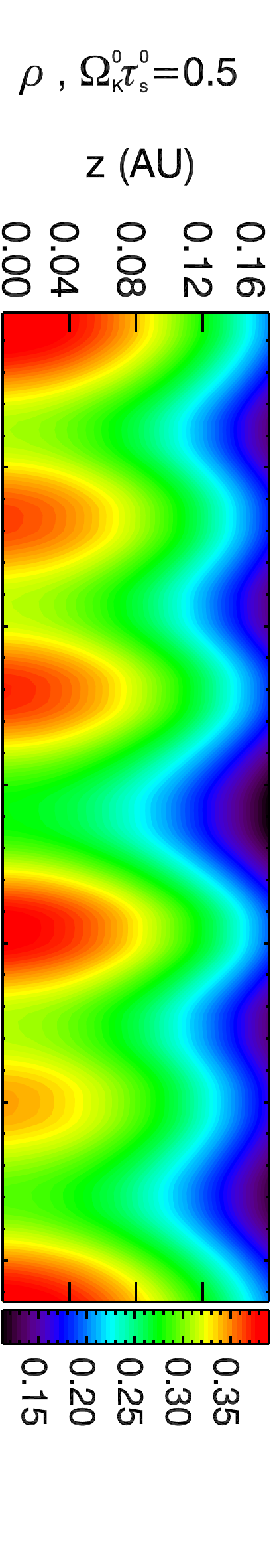}   
   \includegraphics[height=0.9\textwidth,trim=2.6cm 0cm 3.2cm 2cm,clip=true,angle=90]{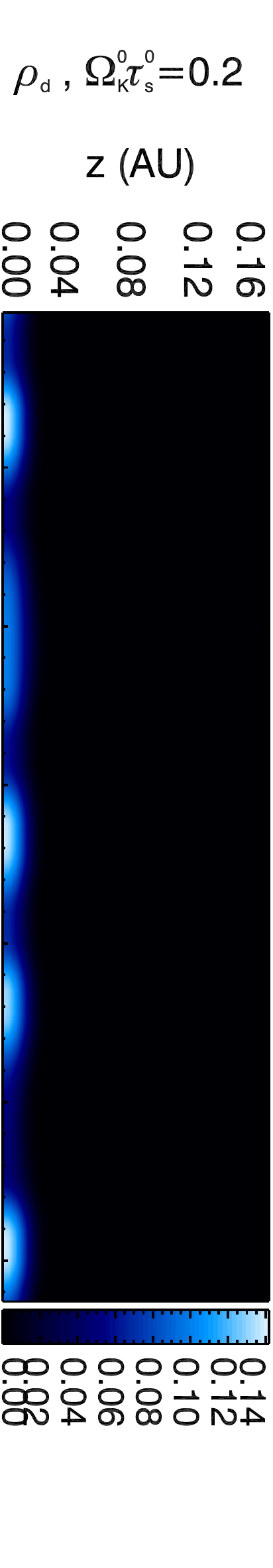}
   \includegraphics[height=0.9\textwidth,trim=2.6cm 0cm 3.2cm 2cm,clip=true,angle=90]{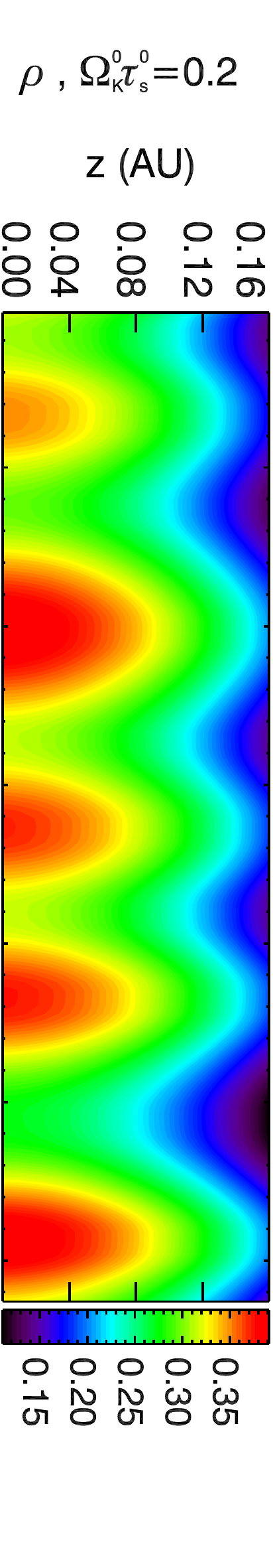}  
   \includegraphics[height=0.9\textwidth,trim=2.6cm 0cm 3.2cm 2cm,clip=true,angle=90]{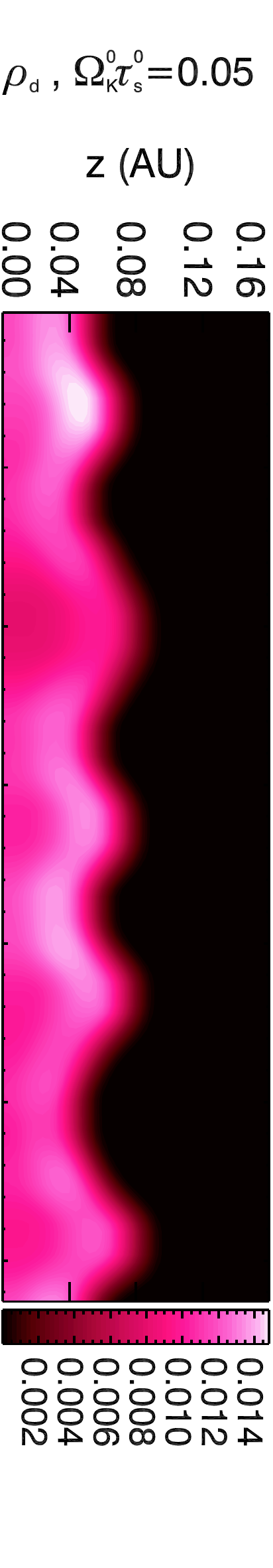}
   \includegraphics[height=0.9\textwidth,trim=2.6cm 0cm 3.2cm 2cm,clip=true,angle=90]{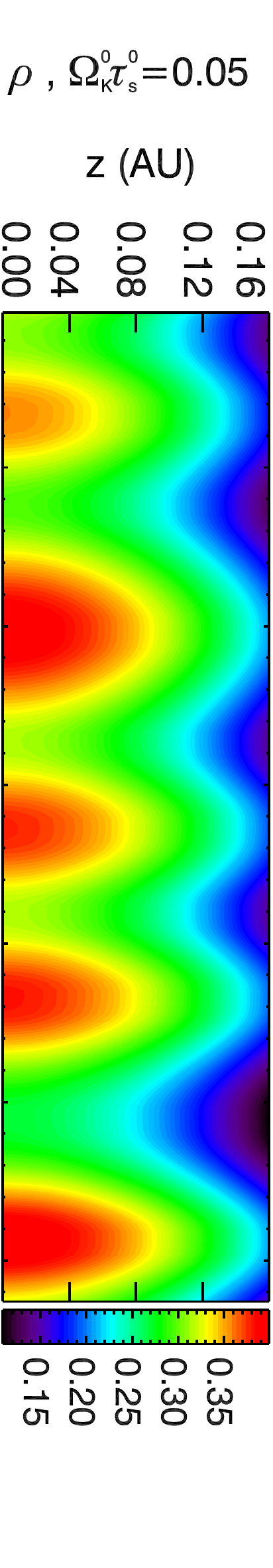} 
   \includegraphics[height=0.9\textwidth,trim=2.6cm 0cm 3.2cm 2cm,clip=true,angle=90]{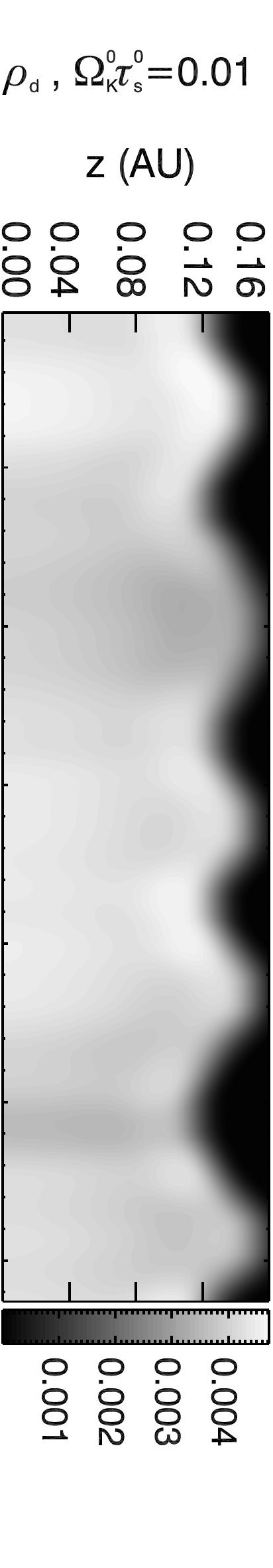}   
   \includegraphics[height=0.9\textwidth,trim=0.5cm 0cm 3.2cm 2cm,clip=true,angle=90]{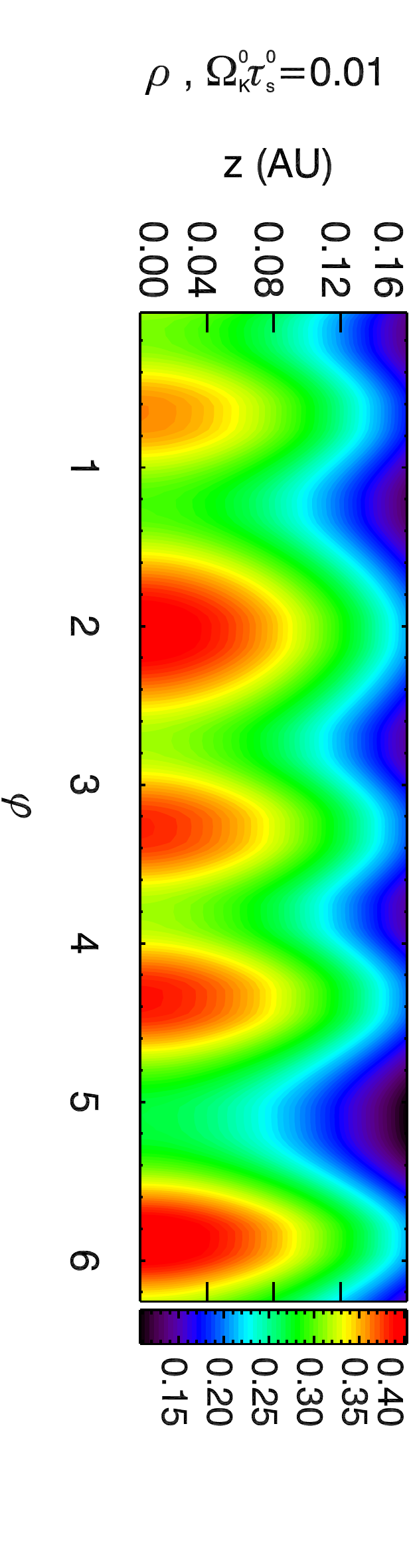}   
    \caption{Vertical density profiles in the vortices ($r/r_0=3$) after $3$ rotations for three different population of dust: $\Omega_K^0\tau_s^0=0.2,0.05$, $0.01$ and for the gas in each of theses cases. As the ranges of density vary largely between populations, different colour tables are used to avoid misunderstandings. The position of the vortices is not the same in the $5cm$ grain simulation due to the back-reaction of the dust on the gas as explained in section \ref{sec:back}.
              }
    \label{Fig:rhovert}
\end{figure*}

One particularly interesting result of these simulations deals with the vertical stratification of the dust.
This is shown in \mbox{Fig.~\ref{Fig:rhovert}}, where the dust density in the $(\varphi,z)$ plane at the radius of the vortices is shown for different dust sizes. 
At the end of the simulation, the largest particles have settled and are concentrated in a very thin region (as can be seen on the upper plots). 
Due to the presence of the vortices, the larger grains ($\ge 2cm$) do not form a continuous disk but a few piles with very high density corresponding to the positions of the five anticyclonic vortices. 

\begin{figure}
	\centering
   \includegraphics[height=9cm,trim=0.5cm 1.cm 1.2cm 4cm,clip=true,angle=90]{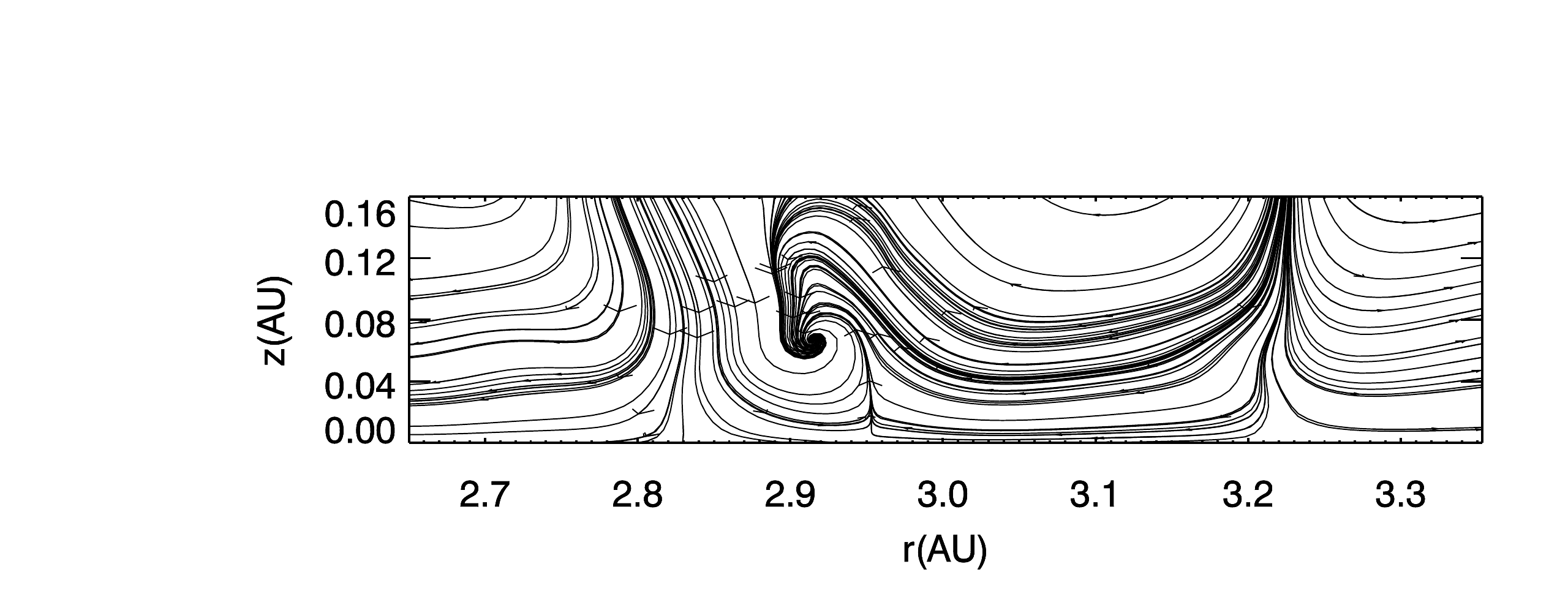}
    \caption{Gas streamline in a vertical plane situated in an anticyclonic vortex. It shows the vertical displacements inside the vortices. The full 3D streamlines are shown in \citet{MKC12}
    \label{Fig:streamrz}
              }    
\end{figure}

The smaller grains ($<2cm$) show a totally different vertical structure with a higher disk height, which is directly related to the dust size, but the same asymmetry is obtained with a thicker disk in the anticyclonic vortices. 
Contrary to what could be expected for settling grains, the density of small particles is higher in the \emph{upper} region than in the mid-plane (see for instance the $5mm$ grains). 
This counter-intuitive result is related to the vertical displacement of the gas inside the vortices as plotted in Fig.~\ref{Fig:streamrz} and detailed in \citet{MKC12}. 
This 3D structure of the vortices' velocity streamlines is responsible for the \emph{positive vertical velocity of the small grains} that are then lifted toward the upper region. 
The vertical profile of dust density in the centre of an anticyclonic vortex is presented in \mbox{Fig.~\ref{Fig:vert}}.
Even with the log-log scale, the increase of density in the upper region is visible for the smaller grains (\emph{e.g.} $\Omega_K^0\tau_s^0$=0.05). 
As the gas density is decreasing with height, the highest dust-to-gas ratio is then obtained in the upper region of the dust disk. 

For each dust population, the gas density in the same vertical plane is also plotted. 
It should be noted that the vertical structure of the gas is the same for all grain sizes. 
Even the $5cm$ grains with a high dust-to-gas ratio do not modify significantly the vertical extent of the gas. 
As obtained in our previous simulations, the height of the gas is not axisymmetric, with a thicker disk in the anticyclonic vortices.

\subsection{Back-reaction of the dust on the gas}\label{sec:back}

\begin{figure}
	\centering
   \includegraphics[height=15cm,trim=3.2cm 1cm 0.5cm 0cm,clip=true]{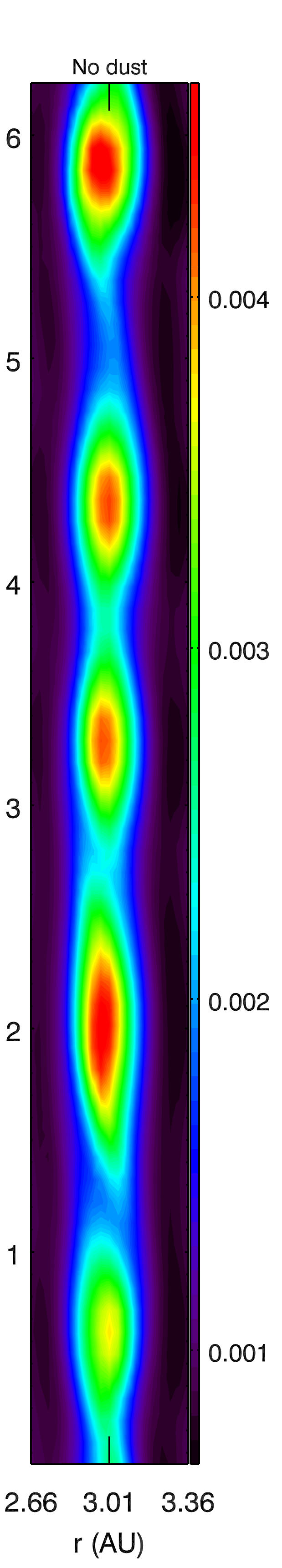}
   \includegraphics[height=15cm,trim=3.2cm 1cm 0.5cm 0cm,clip=true]{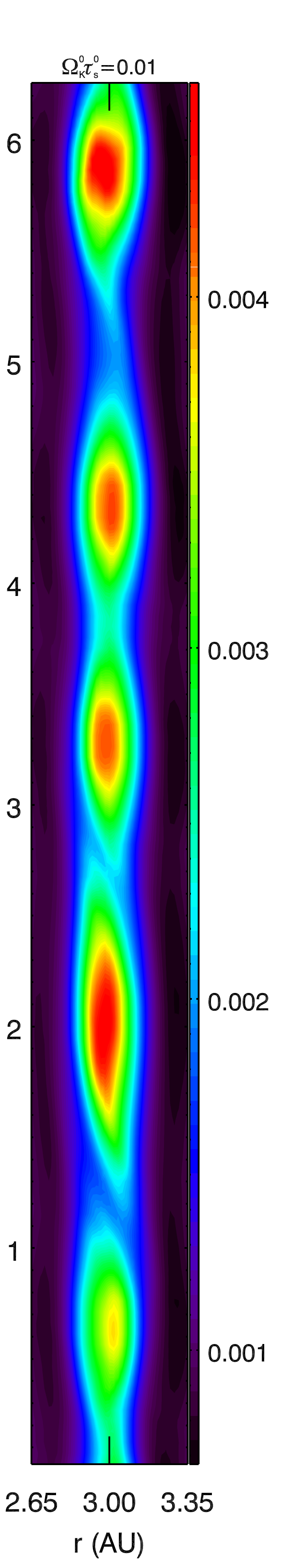}	
   \includegraphics[height=15cm,trim=3.2cm 1cm 0.5cm 0cm,clip=true]{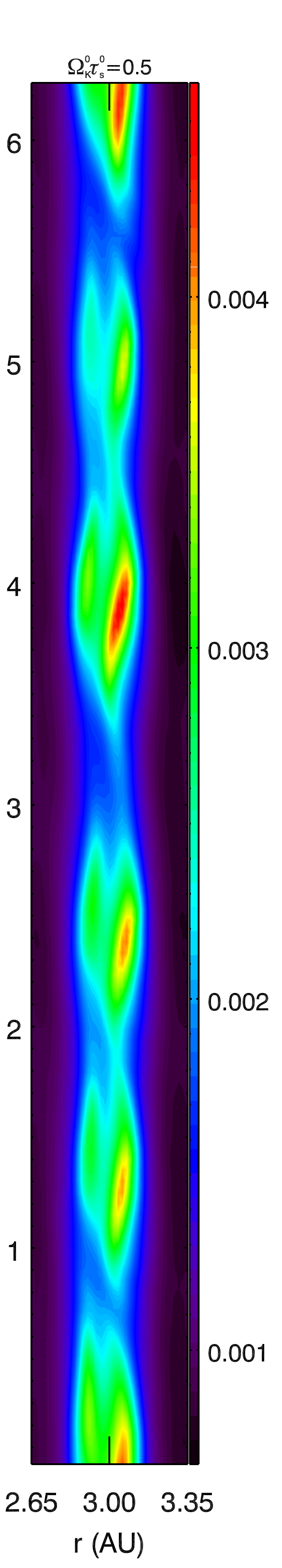}
    \caption{Inverse of gas vortensity at $t/T_B=3$ for the simulation without any dust populations, and the two cases \mbox{$\Omega_K^0\tau_s^0=0.01$} and \mbox{$\Omega_K^0\tau_s^0=0.5$}. In the latter case, there is a modification of the structure of the vortices due to the back-reaction of dust on the gas.
              }  
    \label{Fig:vort}              
\end{figure}

The simulations include the back-reaction of the dust on the gas. 
For most of the dust populations, this back-reaction is negligible due to a low dust-to-gas ratio. 
But for the $5cm$ grains, the dust-to-gas ratio is of the order of unity in the anticyclonic vortices, and the back-reaction modifies the evolution of the gas: the dust drags the gas.
In the absence of the drag force, the dust rotates in keplerian rotation whereas the gas is sub-keplerian in a negative pressure gradient. 
When the dust starts to affect the gas, it accelerates the rotation of the gas and the vortices amplitude decreases. 
Whereas \citet{JAB04} observed a stretching of the vortex by differential rotation, our simulations with self-forming vortices show a split in two different vortices as shown in \mbox{Fig.~\ref{Fig:vort}}. 
This evolution appears only for the $5cm$ grains, no such behaviour is observed with smaller grains. 
This is correlated with the high dust-to-gas ratio reached with this dust population.
The origin of this splitting of the vortices may be related to the evolution of the RWI under external forcing rather than the heavy core instability \citep{CO10}, as the RWI is characterised by the presence of two Rossby waves, one on each side of the density maximum. 
Gas and dust mutually coupled are expected to trigger the streaming instability \citep{YG05}. 
This instability starts to be relevant when the dust begins to drag the gas and the simulations are performed up to this threshold.
 
In the absence of dust, the vortices are rotating with nearly the azimuthal velocity of the gas at the density bump ($r=3AU$) which is keplerian, so the vortices do approximatively $3$ rotations over the simulation. 
See \citet{MYL12} for the calculation of the vortices's velocity when there is no back-reaction of the dust on the gas. 
As the dust accelerates the rotation of the gas, the vortex frequency is no longer that of the wave amplified by the RWI. The vortices, when not sustained by the instability, begin to decay.
Of course, this applies to those vortices formed without dust and whose frequency is determined by the 'classical' RWI. 
For this reason, a dusty RWI should be investigated but this is beyond the scope of this paper.

\section{Summary and outlooks}\label{sec:discussion}
 
We have studied the concentration of dust particles in 3D vortices. 
To our knowledge this is the first time the dust-trapping mechanism has been explored in stable three-dimensional Rossby vortices. 
We have first done a simulation of the self-consistent formation the vortices by the Rossby wave instability before including the dust particles. 
An important difference with the previous studies using analytical vortices (\emph{e.g.} \citealt{K81}) is the presence of a vertical velocity in the inner part of the vortices. 
We have presented the dust-trapping properties of the 3D Rossby vortices. 
This mechanism is very efficient when the dust is only partially coupled to the gas ($\Omega_K^0\tau_s^0=0.5$), and a high dust density is reached in the mid-plane. 
The estimation of the dust mass concentrated in the vortices gives a value of approximately the mass of Mars in a sphere of radius $0.1AU$ with a higher density reached in the centre. 

Those particles more coupled to the gas show a larger density in the upper region of the disk. 
For these intermediate size grains ($mm$ to $cm$ sizes), there is a competition between sedimentation toward the mid-plane and lifting toward the surface by the vertical velocity of the vortices. 
This high dust density in the upper region is of particular interest in the context of the forthcoming observation of protoplanetary disks but this needs to be confirmed by the use of a radiative transfer code with a full 3D approach. 
The results may differ from those obtained with a razor thin disk approach \citep{WK02,RJS11}.

 This mechanism is accordingly more convincing since its efficiency is the highest for the fastest drifting solids, namely when the stopping time is of the order of unity.
  Future work should study the growth of the instability in the presence of dust to understand how the dust modifies the instability. 
  Moreover a simulation with multiple dust sizes in a multi-fluid simulation is necessary to understand how the small dust is concentrated if the vortices start to be accelerated by the larger particles. 
  Furthermore, we have considered a gaussian pressure bump without considering its formation process, which would give the proper shape of the bump, and then the characteristics, including amplitude, of the RWI. 
  A global study, including accretion processes in the disk, is still needed to give the amplitude of the bump, the consequent number of vortices and then the amount of dust concentrated in such vortices. 
  An important step in this direction was done by \citet{LM12}.

Finally, in this paper we associated a stopping time with a dust size and fixed density, but the opposite approach can also be used to study the behaviour of dust grains of the same size but different composition.

\begin{acknowledgements}
This work was partially supported by the Swiss National Science Foundation. We thank M. Houck for his useful comments that helped to improve the manuscript.
\end{acknowledgements}

\end{document}